\documentclass[11pt]{article}

\usepackage{amsmath}
\usepackage{amssymb}
\usepackage{latexsym}
\usepackage{bbm}
\usepackage{graphics}
\usepackage{epsfig}
\usepackage{color}
\if@twoside\oddsidemargin25mm
\else
\renewcommand{\d}{\mathrm{d}}

\newcommand{\Ra}{\Rightarrow}

\DeclareMathSymbol{\mg}{\mathrel}{symbols}{"1D}

\newcommand{\ga}{\alpha}
\newcommand{\gb}{\beta}
\renewcommand{\gg}{\gamma}
\newcommand{\gd}{\delta}
\renewcommand{\ge}{\epsilon}

\newcommand{\gf}{\phi}

\newcommand{\gm}{\mu}
\newcommand{\gn}{\nu}

\newcommand{\gr}{\rho}

\newcommand{\gs}{\sigma}

\newcommand{\go}{\omega}
\newcommand{\gz}{\zeta}
\newcommand{\gp}{\pi}
\newcommand{\gps}{\psi}
\newcommand{\get}{\eta}

\newcommand{\gG}{\Gamma}

\newcommand{\gL}{\Lambda}

\newcommand{\gO}{\Omega}

\newcommand{\cA}{{\cal A}}

\newcommand{\cF}{{\cal F}}

\newcommand{\cH}{{\cal H}}

\newcommand{\cO}{{\cal O}}
\newcommand{\cP}{{\cal P}}

\newcommand{\cR}{{\cal R}}

\newcommand{\cT}{{\cal T}}

\newcommand{\Tr}{\mbox{Tr}}
\newcommand{\tr}{\text{tr}}

\newcommand{\Id}{\mathbbm{1}}
\newcommand{\slashed}{\hspace{-1.1ex}/}
\newcommand{\Slashed}{\hspace{-1.4ex}/\hspace{.2ex}}

\newcommand{\ra}{\rightarrow}

\newcommand{\der}{\partial}

\newcommand{\inv}{^{\mbox{-}1}}
\newcommand{\labl}[1]{\label{#1}}
\newcommand{\beq}{\begin{equation}}
\newcommand{\eeq}{\end{equation}}
\newcommand{\barr}{\begin{array}}
\newcommand{\earr}{\end{array}}
\newcommand{\equ}[1]{\begin{gather} #1 \end{gather}}
\newcommand{\equa}[1]{\begin{align} #1 \end{align}}

\newcommand{\arry}[2]{\begin{array}{#1} #2 \end{array}}

\newcommand{\pmtrx}[1]{\begin{pmatrix} #1 \end{pmatrix}}

\newcommand{\non}{\nonumber}
\newcounter{oldcounter}

\newcommand{\fZ}{\mathfrak{ Z}}
\newcommand{\bgps}{{\bar\psi}}

\newcommand{\Intr}{\mathbb{Z}}
\newcommand{\Cplx}{\mathbb{C}}
\newcommand{\Real}{\mathbb{R}}
\newcommand{\Mnfd}{\mathbb{M}}
\newcommand{\Tors}{\mathbb{T}}
\newcommand{\Fixd}{\mathbb{T}{}^{\rm fix}}

\newcommand{\Fibr}{\mathbb{F}}

\newcommand{\ba}[2]{\[\begin{array}{#2}\label{#1}}
\newcommand{\ea}{\end{array}\]}
\newcommand{\be}{\begin{equation}}
\newcommand{\ee}{\end{equation}}
\newcommand{\bea}{\begin{eqnarray}}
\newcommand{\eea}{\end{eqnarray}}

\newcommand{\E}[1]{\mathrm{E_{#1}}}
\newcommand{\GG}{\mathrm{G_{2}}}
\newcommand{\U}[1]{\mathrm{U(#1)}}
\newcommand{\SU}[1]{\mathrm{SU(#1)}}
\newcommand{\SO}[1]{\mathrm{SO(#1)}}

\newcommand{\parl}{{\,{\rotatebox{90}{\small $=$}}}}

\newcommand{\rep}[1]{\mathbf{#1}}

\begin{document}

\begin{flushright}
hep-th/0305139
\\
UVIC-TH/04-03
\end{flushright}
\vskip 2 cm
\begin{center}
{\Large {\bf 
Traces on orbifolds: Anomalies and one--loop amplitudes 
} 
}
\\[0pt]

\bigskip
\bigskip {\large
{\bf Stefan Groot Nibbelink}\footnote{
{{ {\ {\ {\ E-mail: grootnib@uvic.ca}}}}}}}, 
\bigskip 
\\[0pt]
\vspace{0.23cm}
{\it 
University of Victoria,  Dept.\ of Physics \& Astronomy, \\
PO Box 3055 STN CSC, Victoria, BC, V8W 3P6 Canada.\\
(CITA National Fellow)\\
}

\bigskip
\vspace{1.4cm} 
\end{center}
\subsection*{\centering Abstract}

In the recent literature one can find calculations of various
one--loop amplitudes, like anomalies, tadpoles and vacuum energies, on
specific types of orbifolds, like $S^1/\Intr_2$. 
This work aims to give a general description of such one--loop
computations for a large class of orbifold models. In order to achieve
a high degree of generality, we formulate these calculations as
evaluations of traces of operators over orbifold Hilbert spaces. 
We find that in general the result is expressed as a sum of traces
over hyper surfaces with local projections, and the derivatives
perpendicular to these hyper surfaces are rescaled. 
These local projectors naturally takes into account possible
non--periodic boundary conditions. 
As the examples $T^6/\Intr_4$ and $T^4/D_4$ illustrate, the
methods can be applied to non--prime as well as non--Abelian
orbifolds.

\newpage

\section{Introduction}
\labl{sc:intro}

In both string theory and field theories of extra dimensions one
often considers compactifications on orbifolds. This can either be on
their own right, or as approximations of more complicated smooth spaces,
like Calabi--Yau or $\GG$ manifolds, for example. The simplest orbifold
$S^1/\Intr_2$ has been studied in the context of the eleven dimensional
supergravity limit of M--theory \cite{Horava:1996ma,Horava:1996qa}.  
This orbifold also received a lot of attention in five dimensional 
(GUT) models with (broken) supersymmetric 
\cite{Antoniadis:1990ew,Delgado:1998qr,Barbieri:2000vh,Barbieri:2001yz,
Hall:2001pg,Hall:2001xb,Hall:2002ci,Delgado:2001si,Hebecker:2002vm,
Hebecker:2001wq,Kawamura:2000ev,Altarelli:2001qj,Hebecker:2001jb}, 
and has even been studied at two loop \cite{Delgado:2001xr}. 
Clearly, since $S^1/\Intr_2$ can not be blown up to a smooth manifold,
all examples based on this orbifold cannot be related to theories
on smooth spaces. On the contrary, four and six dimensional orbifolds, 
like $T^4/\Intr_2$ or $T^6/\Intr_3$, can be resolved to give rise to
smooth manifolds. All resolutions of four dimensional orbifolds with
discrete subgroup of $\SU{2}$ result in a (topologically) single
manifold, called $K3$. For the six 
dimensional orbifolds, the possible resolutions are classified by two
Hodge numbers. Somewhat surprising, in some sense both string and
field theory on orbifolds do not seem to care about whether the
orbifolds have resolutions or not: Calculation of physical quantities
on either type of orbifolds proceed in an identical fashion. 
Let us mention a few important computations on orbifolds that are often
considered.

Recently there have been substantial investigations to the profile 
of gauge anomalies on orbifolds.
In the context of string theory there have been many investigations to
anomalies of zero modes \cite{Candelas:1985en,Dixon:1986jc}.  
Horava and Witten were the first to argue, that anomalies on
$S^1/\Intr_2$ would distribute equally at both fixed points
\cite{Horava:1996ma,Horava:1996qa}, which lead to the discovery of
heterotic M--theory. A direct calculation of the shape of the anomaly
over $S^1/\Intr_2$ has been performed by Arkani--Hamed, Cohen and
Georgi with a gauge field and a fermion \cite{Arkani-Hamed:2001is},
confirming the Horava--Witten expectation in a five dimensional
setting. After that various groups computed the structure of the
anomaly on the orbifold $S^1/\Intr_2\times\Intr_2'$, see
\cite{Scrucca:2001eb,Pilo:2002hu,GrootNibbelink:2002qp}. 
For an investigation of anomalies in a seven dimensional setting
compactified on $S^1/\Intr_2$ see refs.\ 
\cite{Gherghetta:2002xf,Gherghetta:2002nq}. 
Investigations of the shape of anomalies in more than one extra
dimensions have also been performed: In ref.\ \cite{Gmeiner:2002es}  
the gaugino anomalies in heterotic string theory on $T^6/\Intr_3$ with
Wilson lines were computed. Anomalies of a six dimensional model on
$T^2/\Intr_2^3$ can be found in ref.\ \cite{Asaka:2002my}. Very
recently, an investigation of anomalies on a more general class of
orbifolds has been presented in ref.\ \cite{vonGersdorff:2003dt}.

Anomaly investigations are very important, as they may provide us with
important quantum consistency constraints. However, also other quantum
corrections can be vital to gain a more complete understanding of the
physics described by a given theory. Let us mention a few of such
effects, that have been considered in the context of extra
dimensions. To investigate stability issues of a higher dimensional
theory from the four dimensional point of view, the effective zero
mode potential may proof an important tool
\cite{Ponton:2001hq,Kanti:2002zr,Hofmann:2000cj,Bucci:2003fk,
Garriga:2000jb,Flachi:2003bb,Albrecht:2001cp}. 
This potential is obtained by integrating out all Kaluza--Klein modes
from the theory. In models where also gravitational effects are
considered, one can determine whether the extra dimensions are
stabilized and that late time four dimensional cosmology is a
possibility, for example.

A different type of stability issue is concerned with the degree of
divergence of the fundamental parameters in the classical
Lagrangian. Most notably, a quadratically divergent Higgs (scalar)
mass parameter leads to the well--know hierarchy problem. In 
supersymmetric theories this divergent scalar mass parameter can be
reformulated as a Fayet--Iliopoulos tadpole \cite{Fayet:1974jb} 
for an auxiliary field component of a supersymmetric (vector)
multiplet. In exact supersymmetry, this term is only renormalized at
the one--loop level \cite{Fischler:1981zk}. This essentially four
dimensional discussion has been lifted to five dimensional 
orbifold theories, showing that these tadpoles can arise on the
boundaries of orbifolds, like $S^1/\Intr_2$ and
$S^1/\Intr_2\times \Intr_2'$ 
\cite{Ghilencea:2001bw,Barbieri:2001cz,Barbieri:2002ic,
GrootNibbelink:2002wv,GrootNibbelink:2002qp}, and their consequences
have been studied in  \cite{Marti:2002ar,Abe:2002ps}. 
Also in higher dimensional cases such
tadpoles can be discussed: In six dimensional models this has recently
been considered in \cite{vonGersdorff:2002us,Csaki:2002ur}.

Of ten dimensional super Yang--Mills, no auxiliary field formulation
exists. However, as has been shown in ref.\
\cite{Marcus:1983wb,GrootNibbelink:2003gb}, 
one can introduce supersymmetric
auxiliary field components w.r.t.\ one of the 4 supersymmetries in
four dimensions. On the fixed points of the orbifold $T^6/\Intr_3$
these auxiliary fields can develop similar Fayet--Iliopoulos tadpoles 
for the local anomalous $\U{1}$s. (The tadpoles of zero mode anomalous
$\U{1}$s have been discussed in detail in the past in the context of string
theory  \cite{ Dine:1987xk,Atick:1987gy,Dine:1987gj,Kobayashi:1997pb, 
Poppitz:1998dj})

As can been seen from these examples given above, there has not yet
been a complete and coherent discussion of all such quantum
computations: Only some particular types of orbifolds, and quantum
amplitudes have been considered. This article aims to give a unified
description of all possible one--loop quantum amplitudes on arbitrary
(flat) orbifolds in any given number of dimensions. These orbifolds
may be compact or non--compact (like $\Real/\Intr_2$, for example) or
even non--Abelian. Taking the orbifold group isomorphic to a finite
subgroup of $\SU{2}$, orbifolds are obtained that have so--called ADE 
singularities. In addition, in the compact directions of the
orbifold one may introduce Wilson lines or Scherk--Schwarz
(super)symmetry breaking
\cite{Scherk:1979ta,Hosotani:1989bm}. (Discrete Wilson lines were
first considered in the context of string theory in 
\cite{Ibanez:1988pj,Font:1988tp}.)

In order to arrive at a sufficiently general discussion of all these
different quantum computations for various orbifold theories, it is
convenient to employ a somewhat more abstract description, in terms of
Hilbert spaces associated to bundles over orbifolds. Let us recall the
connection between these mathematical concepts and the field theory. 
Any field defines a function of spacetime to a complex vector
space. If the spacetime is topologically non--trivial, the field is
described by a (local) section of the corresponding complex fiber
bundle. The inner product on this fiber can be used to define an inner
product on the space of all such sections, turning it into a Hilbert
space. One--loop amplitudes can then be formulated as the traces of
operators over this Hilbert space. This basic strategy allows us to
arrive at an unified description of the computation of all such
one--loop quantities on orbifolds.

\subsection*{Paper organization}

Section \ref{sc:bundlOrbi} is devoted to provide the mathematical
basis for this paper. Subsection \ref{sc:geom} discusses the geometry
of flat orbifolds, in terms of the properties of a torus lattice $\gG$
and an orbifold group $G$. The later part of that subsection
identifies the orbifold fixed points, and describes their properties
in some detail. The following subsection describes bundles on
orbifolds in terms of homomorphisms of the torus lattice and the
orbifold group. The Hilbert spaces associated with these bundles are
introduced. Subsection \ref{sc:HilbertIsos} describes a useful
isomorphism between Hilbert spaces of periodic fields and fields that
are periodic up to the orbifold group homomorphism. In addition, a
projection operator on the torus Hilbert space is defined to obtain
states, that descent down onto the orbifold.

Section \ref{sc:main} is the main part of this work. In subsection
\ref{sc:OrbiTrace} traces on the various Hilbert spaces are
defined. Next, the trace on a general orbifold is evaluated; the
details of this calculations have been collected in appendix
\ref{sc:traces}. The next two subsections discuss a couple of
important applications of the results obtained in this work: the
computation of anomalies, tadpoles and vacuum energies on orbifolds.

The material in these two sections has been presented in a rather
abstract fashion, therefore it might be helpful to the reader to see
how the various concepts can be applied to concrete
examples of orbifolds. In section \ref{sc:Examples} we consider three
orbifolds in various dimensions: First, in
subsection \ref{sc:S1Z2} we calculate the trace of a general operator
on the well--studied orbifold $S^1/\Intr_2$ as familiar
illustration. In the following subsection we focus on the orbifold
$T^6/\Intr_4$ which has an orbifold group that has a subgroup, namely 
$\Intr_2$.  In subsection \ref{sc:T4D4} we give an example of a 
non--Abelian orbifold $T^4/D_4$, where $D_4$ is the 
dihedral group with eight elements.

Finally, in section \ref{sc:concl} we summarize the main conclusions
of this work. In appendix \ref{sc:scalar} we expose a couple
useful properties of scalar mode functions on tori.

\section{Hilbert spaces associated with bundles over orbifolds}
\labl{sc:bundlOrbi}

In this section we develop some general material to describe field
theories on a large class of orbifolds. The discussion here is
necessarily a bit abstract and could be some what difficult to follow at
first reading. Therefore, we encourage the reader to simultaneously
consider the examples provided in section \ref{sc:Examples}. These
examples have been chosen both to be familiar to the reader (the
$S^1/\Intr_2$ example) as well as to illustrate all subtleties that
arise for non--prime and non--Abelian orbifolds.

\subsection{Orbifold geometry}
\labl{sc:geom}

We begin by defining the orbifold $\Tors/G$ which is sufficiently
general, to allow us to describe the various types of orbifolds
referred to in the introduction. The space $\Tors = \Real^{1,d-1}/\gG$ is
defined as a quotient, using an integral lattice $\gG \cong \Intr^n$ in
$d$--dimensional Minkowski space $\Real^{1,d-1}$. In the context of
compactification one considers the case where $\gG$ act on the spatial
part of the Minkowski space, then we obtain $d-n$ dimensional
Minkowski space times an $n$ dimensional torus: 
$\Tors = \Real^{1,d-n-1} \otimes T^n$ with $n <d$. If $\gG$ has one
basis vector in the time direction, we can use $\Tors$ to perform
finite temperature calculations. 
The inner product $x^T\get \, y$ for $x, y \in \Real^{1,d-1}$ is
defined in terms of the diagonal matrix 
$\get = \text{diag}(-1, 1,\dots, 1)$  and $x^T$ denotes the transposed
of the vector $x$. Let $G$ be a finite group, that acts on
$\Real^{1,d-1}$, and preserves the Minkowskian inner product and the
orientation of $\Real^{1,d-1}$; we take $G \subset SO(1,d-1)$. 
In addition, it 
has to be compatible with the lattice $\gG$ that defines the torus, i.e.\ for
all $g \in G$ we have $g \gG = \gG$. In general the group $G$ does not
act freely on $\Tors$; there may be subspaces of $\Tors$ which are
fixed elements of the orbifold group. It should be stressed that
elements of $G$ may act in both the compact and non--compact 
directions of $\Tors$. To mention a few concrete examples of orbifolds
that can be treated using this formalism: 
$\Real^{1,3}\otimes\Real/\Intr_2$, 
$\Real^{1,3}\otimes(T^2 \otimes \Cplx)/\Intr_4$ and the examples
considered in section \ref{sc:Examples}.

Each element $g \in G$ generates
an Abelian subgroup $\langle g \rangle = \{ g^k \,|\, 0 \leq k<|g|\}$,
which is isomorphic to $\Intr_{|g|}$. Here $|g|$ is the order of $g$, i.e.\
the smallest number $n$ such that $g^{n} = 1$.  
We define an operator $P{}_g^\parl$ for each $g \in G$, that  projects
on the subspace $\Tors{}^\parl_g$ on which $g$ acts as the
identity. This definition, and many others that follow below, are all
assumed to be defined on the covering space $\Real^{1,d-1}$. After
that, it is not difficult to apply these definitions to $\Tors$ as
well.  This projection operator has the following properties 
\equ{
P{}^\parl_{g} = \frac 1{|g|} \sum_{k =1}^{|g|} g^k, 
\qquad 
P{}_g^\parl = (P{}_g^\parl)^2 = g \, P{}_g^\parl = P{}_g^\parl \,g 
= P{}^\parl_{g\inv}.
\labl{gProj}
}
In addition, we have that $  = P{}^\parl_{g^p} = P{}^\parl_g$ 
if $p$ and $|g|$ are relatively prime, i.e.\ $\gcd(p, |g|) =1$,  
since then $g^p$ and $g$ generate the same subgroup.
Moreover, we define $P{}^\perp_g = 1 - P{}^\parl_g$. 
Using these projectors, we can decompose the space $\Tors$ as 
\equ{
\Tors = \Tors{}^\parl_g \otimes \Tors{}^\perp_{\,g}, 
\qquad 
\Tors{}_g^\parl = P{}_g^\parl\,  \Tors,
\qquad 
\Tors{}_{\,g}^\perp = P{}_{\,g}^\perp\,  \Tors. 
\labl{DirectProd}
}
The traces $d{}_g^\parl = \tr~P{}_g^\parl$ and 
$d{}_g^\perp = \tr~P{}_g^\perp = d-d{}_g^\parl$ give the real
dimensionality of $\Tors{}_g^\parl$ and $\Tors{}_{\,g}^\perp$,
respectively. By construction $\langle g \rangle$ acts trivially on
$\Tors_g^\parl$, while $g$ does not act freely on $\Tors_{\,g}^\perp$. 
We define the fixed space of $g$ in $\Tors$ as 
\equ{
\Fixd_g = \{x \in \Tors ~|~ (1-g) x \in \gG\}. 
}
The codimension of the space fixed by $g$ is equal to
$d{}_{\,g}^\perp$; i.e.\  $\Fixd_{\,g} \cap \Tors{}_{\,g}^\perp$ gives
exactly the set of fixed points of $g$ in $\Tors{}_{\,g}^\perp$.

In order to obtain unique definitions of these fixed points in
$\Tors{}_{\,g}^\perp$ in the covering space $\Real^{1,d-1}$,  we 
make the following definition of the fundamental domain of the torus
$\Tors$ in this covering space:  
\(
\cF(\Tors) = \{ x^i e_i + y^a \tilde e_a ~|~ 0 \leq x^i < 1, ~
y^a \in \Real  \}
\) 
where $e^i$ are $n$ basis vectors of $\gG$ and $\tilde e_a$ are the
additional basis vectors pointing in the non--compact directions 
to form a basis for $\Real^{1,d-1}$. The
fixed points $\fZ_g^s \in \cF(\Tors{}^\perp_{\,g})$ of 
$\Tors{}^\perp_{\, g}/\langle g \rangle$ 
are chosen such that they have smallest distance to the origin in 
the fundamental domain $\cF(\Tors)$. This description of the fixed 
points 
\equ{
g\, \fZ{}_g^s - \fZ{}_g^s = v{}_{g}^{s} \in \gG 
\labl{vgsDef}
}
defines the vectors $v{}_{g}^{s}$ uniquely w.r.t.\ this fundamental 
domain. The fixed space of $g$ can be decomposed as  
\equ{
\Fixd_{\,g} = \sum_{s} \Fixd_{g\,s} + \gG, 
\qquad 
\Fixd_{g\,s} = \fZ{}_g^{s} + \Tors{}_g^\parl \cong 
\Tors{}_g^\parl. 
\labl{FixdSpace}
}
We define the delta function on $\Tors$ as  
\(
\gd_\Tors(x) = \gd(x-\gG),
\)
where $\gd(x)$ is the delta function on $\Real^{1,d-1}$. Because of
the direct product structure \eqref{DirectProd} of $\Tors$ w.r.t.\ 
$g \in G$, the torus delta function factorizes as  
\(
\gd_{\Tors}(x) = \gd{}_g^\parl(x) \, \gd{}_{\,g}^\perp(x). 
\) 
Here 
\(
\gd{}_{\,g}^{\parl,\perp}(x) =
\gd{}_{\,g}^{\parl,\perp}(P{}_{\,g}^{\parl,\perp}x)
\)
denote the delta function on $\Tors_{\,g}^{\parl,\perp}$. (We assume
for any other object with ${}_{\,g}^{\parl,\perp}$ that its coordinate
dependence has been restricted to $\Tors_{\,g}^{\parl,\perp}$.) 
The delta function $\gd_{\,g}^\perp$ has the important property:   
\equ{
\gd_{\,g}^\perp( x-g\, x ) = 
\frac 1{|\det^\perp_{\,g} (1\!\!-\!g)|} \sum_{s} 
\gd_g^\perp(x - \fZ_{g}^{s}).
\labl{FixedDelta} 
}
Here $\det{}^\perp_{\,g}(1-g)$ denotes the determinant on 
$\Tors{}^\perp_{\,g}$, and for that reason it is non--vanishing. 
In particular, we set it equal to unity for $g=1$.

For any function $F$ we define the integral over $\Fixd_{\, g}$ as 
\equ{
\int_{\Fixd_{\,g}} \d x\, F(x) = 
\sum_{s}  \int_{\Tors} \d x\, 
F(x^\parl_g +  \fZ_{g}^{s}) \, 
\gd_{\,g}^\perp(x - \fZ_{g}^{s}). 
\labl{IntFixdDelta}
}

Many of the properties listed above only depend on the conjugacy class
$(g) = \{ h\inv g h ~|~ h \in G \}$ in the group $G$, but not on the
particular element $g$ in the conjugacy class. To show this, we
observe that under conjugation we have the following identities  
\equ{
\arry{c}{
P{}^{\parl,\perp}_{h g h\inv} = h \, P{}^{\parl,\perp}_{\,g} \, h\inv, 
\qquad 
\Tors^{\parl,\perp}_{h g h\inv} = h \, \Tors^{\parl,\perp}_{\,g}, 
\qquad 
\Fixd_{h g h\inv} = h \, \Fixd_{\,g},
\qquad 
d^{\parl,\perp}_{h g h\inv} = d^{\parl,\perp}_{\,g}, 
\\[2ex]
\fZ^s_{h g h\inv} = h\, \fZ^s_g, 
\qquad
v^s_{h g h\inv} = h\, v^s_g, 
\qquad
\det^\perp_{h g h\inv}(1- h g h\inv) = \det^\perp_{\,g}(1-g), 
}
\labl{Conjugations}
}
using that $h\, \Tors = \Tors$, $h\, \gG = \gG$; and similarly for
the dual vector spaces $\Tors{}^{\parl,\perp *}_g$, defined in
\eqref{DualV}. In each conjugacy class $(g_r)$ we choose
a representative $g_r$. Each element $h \in (g_r)$ can be written as
$h = k g_r k\inv$, with $k$ an element of the coset $G/C(g_r)$, with
the centralizer $C(g) = \{ h \in G \,|\, hgh\inv = g \}$.

With these definitions and relations we can investigate the structure
of the fixed sets $\Fixd_{\, g}$ orbifolded by the action of $G$. 
First of all, because of the conjugation property
\eqref{Conjugations} of $\Fixd_{\,g}$ under the action of $h \in G$ it
follows, that all fixed spaces $\Fixd_{k}$ corresponding to one
conjugacy class, i.e.\  $k \in (g_r)$ are identified.  This leads to
the identity 
\equ{
\Bigl( \bigcup_{\ k \in (g_r)} \!\! \Fixd_{\,k} \Bigr) / G = 
\Fixd_{\, g_r}/C(g_r).
\labl{FixdConjOrbi}
} 
Of course, as soon as the group $G$ is Abelian, each group element is
its own conjugacy class and $C(g) = G$, so that the relation above
becomes trivial. Even though the centralizer $C(g)$ maps $\Fixd_{\,g}$
to itself, it does not necessarily fix all points in $\Fixd_{\,g}$. In
fact, $C(g)$ gives rise to an equivalence relation between the labels
of the fixed points: $s \sim s'$ if there is a 
$h \in C(g)$ such that $h \fZ{}^s_g = \fZ{}^{s'}_g + \gG$. Upon
orbifolding, this leads to the identification of $\Fixd_{g\, s}$ and
$\Fixd_{g\, s'}$. We denote the representatives of the corresponding
equivalence classes by $s_r$, and let  $G{}_{g}^s \subset C(g)$ 
be the subset, which fixes the fixed point $\fZ{}_g^s$ of $g$. 
One should realize that this identification of $\Fixd_{g\,s}$ and
$\Fixd_{g\,s'}$ might be non--trivial. 
To find out how the  identifications are preformed, we
recall that the structure of the fixed space $\Fixd_{g\, s}$ is given
by \eqref{FixdSpace}. Therefore, the identifications are determined by
the mapping   
\equ{
p{}^\parl_g:~ C(g) \ra \text{Diff}(\Fixd_{\, g});
\quad 
p{}^\parl_g(h) = P{}^\parl_g \, h \, P{}^\parl_g = 
h\, P{}^\parl_g = P{}^\parl_g \, h,
\labl{Residual}
}
since $h \in C(g)$. The kernel of this group homomorphism is denoted
by $ \text{Ker}{}_{g}^\parl$. Upon the identifications discussed
above, we find that   
\equ{
\Fixd_{\, g}/C(g) = \sum_{s_r} \Fixd_{g \, s_r}/G{}_g^{s_r},
}
where the sum is over the representatives of the equivalence classes
of fixed points of $g$. This means, that 
$ \Fixd_{g \, s_r}/G{}_g^{s_r}$ represents an orbifold 
of dimension $d^\parl_g$, in general. However, part of the group 
$G{}_g^{s_r}$ may act trivially on $\Fixd_{g \, s_r}$. Since by
definition $G{}_g^{s_r}$ fixes $\fZ_g^{s_r}$, it follows that the
kernel 
\equ{
\text{Ker}{}^\parl_{g\, s_r} = 
\Bigl\{ 
h \in G{}_g^{s_r} \, | \, p{}^\parl_g(h) = P^\parl_g 
\Bigr\} 
= \text{Ker}{}^\parl_{g} \bigcap G{}_g^{s_r}, 
}
characterizes the trivial part of $G{}_g^{s_r}$.Taking all this into
account we finally obtain  
\equ{
\Bigl( \bigcup_{\ h \in (g_r)} \!\! \Fixd_{\,h} \Bigr) / G = 
\sum_{s_r} \Fixd_{g_r\, s_r}/H^{s_r}_{g_r}
\qquad 
H{}^{s_r}_g = G{}^{s_r}_g / \text{Ker}{}^\parl_{g\, s_r}. 
\labl{FixdConjOrbiFinal}
} 
In this formula there is no residual trival action of the orbifold
groups.

\subsection{Bundles of fibers with inner products}
\labl{sc:FiberInner}

Next we give a short description of bundles over the orbifold
$\Tors/G$. To gain an overview of the possible bundles
over this orbifold, it is convenient to start with bundles over the
covering space $\Real^{1,d-1}$. Any bundle over
$\Real^{1,d-1}$ with fiber $\Fibr$ is trivial, i.e.\ it is direct
product $\Real^{1,d-1} \otimes \Fibr$, because the base space 
$\Real^{1,d-1}$ is topologically trivial. We assume, that the fiber
$\Fibr$  is a complex vector space of complex dimension $N$ equipped
with an inner product $\gps^\dag \, \gr\, \gf$ defined for all 
$\gps, \gf \in \Fibr$. Here $\gr$ is an Hermitian matrix. It can be
simply the identity, or the Minkowskian metric $\get$ for the tangent
bundle, or $\gg_0$ for fermions, to name a couple of important
examples.  By enforcing suitable reality conditions, the complex fiber
$\Fibr$ can be turned into a real vector space. On the torus 
$\Tors = \Real^{1,d-1}/\gG$ not all bundles are trivial: since in
general $(\Real^{1,d-1} \otimes \Fibr)/\gG$ is not equal to 
$\Tors \otimes \Fibr$. However, the former is merely notation, 
unless we specify how $\gG$ acts on 
the fiber $\Fibr$. If it acts as the identity, sections are simply
periodic functions $\gps:~ \Tors \ra  \Fibr$, and the bundle is
trivial: $\Tors \otimes \Fibr$. The space $\cH_{\Tors}$ of these
sections is a Hilbert space, w.r.t.\ to the natural inner product   
\equ{
\langle \gps \,|\, \gf \rangle_{\Tors} = \int_{\Tors} \d x\, 
\gps(x)^\dag \, \gr \, \gf(x), 
\quad 
\forall 
\gps, \gf \in \cH_{\Tors}.  
\labl{TorsInner}
}
We can obtain non--trivial bundles by requiring that the section
$\gps$ is only periodic  
\equ{
\gps \in \cH_{\Tors,T}:~~~
\gps(x + v) = T_{v}\, \gps(x),
\qquad 
T_v \in \U{N; \gr},
\qquad 
v \in \gG, 
\labl{TorsBound}
}
up to an 
$\U{N;\gr} = \{ S: \Fibr \ra \Fibr \,|\, S^\dag \gr S = \gr \}$ 
transformation. This requirement ensures that the inner product is
preserved by these transformations. The inner product 
$\langle \gps \,|\, \gf \rangle_{\Tors,T}$ on the corresponding
Hilbert space $\cH_{\Tors,T}$ is defined in the same way as 
\eqref{TorsInner}. The properties of $T_v$ will be developed below.

To define bundles on the orbifold $\Tors/G$ we need to specify how the
group $G$ acts on the fiber, again using sections, we have  
\equ{
\gps \in \cH_{\Tors,T,R}: ~ 
\gps(g\, x) = R_g\, \gps(x),
\qquad 
R_g \in \U{N; \gr},
\qquad 
g \in G.
\labl{BoundConds}
} 
The volume of the orbifold $\Tors/G$ is a factor $1/|G|$ smaller than
that of the torus $\Tors$ (one might say that the torus is made up of
$|G|$ copies of $\Tors/G$). Therefore, the inner product on the
associated orbifold Hilbert space $\cH_{\Tors/G,T,R}$ reads  
\equ{
\langle \gps \,|\, \gf \rangle_{\Tors/G,T,R} = 
\frac 1{|G|}\,  \langle \gps \,|\, \gf \rangle_{\Tors}.
\labl{InnerOrbi}
}
Here we have employed the natural isomorphism between
the Hilbert spaces $\cH_{\Tors,T,R} \ra \cH_{\Tors/G,T,R}$. (We return
to this isomorphism in the next subsection.) This inner product does
not depend on the fundamental domains chosen, since they are invariant
under the identifications induced by $\gG$ and $G$ because of the
unitarity of $T_v$ and $R_g$.

By repeated application of the boundary conditions \eqref{TorsBound}
and \eqref{BoundConds} it follows that, $T_{v}$ and $R_g$ define group
homomorphisms     
\equ{
T_{v + w} = T_v \, T_w, 
\qquad 
R_{g\, h} = R_g \, R_h, 
\qquad 
T_0 = R_1 = 1,
\labl{homom}
}
of $\gG$ and $G$, respectively. Therefore, in particular we have that 
$(R_g)^{|g|} =1$. Since $\gps(x+ v + w - v - w) = \gps(x)$ and   
$\gps(g(g\inv x + v)-g\,v) = \gps(x)$, consistency of these boundary
conditions requires that   
\equ{
[T_{v}, T_{w}] = 0, 
\qquad 
T_{g\, v} = R_g \, T_{v}\, {R_g}\inv.  
\labl{consistency}
}
(Hence, if $R_g$ and $T_v$ commute, one obtains $T_{g\, v} = T_v$.)
Such consistency condition have been discussed in ref.\
\cite{Barbieri:2001dm} in the context of five dimensional models. 
Moreover, we have for any element $g \in G$ that 
\equ{
v \in \gG{}^\perp_{\, g} 
\quad \Ra \quad 
(T_v R_g)^{|g|} = 1, 
\labl{DiscrWilson}
}
with $\gG{}^\perp_{\,g} = P{}^\perp_{\, g} \gG$ since 
$\gps(x + \sum g^k v) = \gps(x)$. This implies that $T_v$ is
quantized. (If $R_g$ and $T_v$ commute, we find that this reduces to 
$T_v{}^{|g|} = 1$. If the order $|g|$ is non--prime additional restrictions may arise, see the examples in sections \ref{sc:T6Z4} and \ref{sc:T4D4}.) 
When the fiber contains a Lie algebra, $T_v$ is
sometimes referred to as a discrete Wilson line.

Finally, we investigate what happens in the bundle at the fixed spaces
of the orbifold. For a point $x$ in the fixed space $\Fixd_{g\,s}$ we
find that 
\(
R_g\, \gps(x) = \gps(g\, x) = \gps(x +v{}_{g}^{s}) 
= T_{v{}_{g}^{s}}\, \gps(x). 
\)
Therefore, at $\Fixd_{g\,s}$ only those states exist, that survive the 
projection
\equ{
R^s_g \, \gps(x) = \gps(x), 
\qquad 
R^s_g = {T_{v{}_{g}^s}}\inv R_g,
\qquad 
x \in \Fixd_{g\, s}. 
\labl{Projgv} 
}

\subsection{Hilbert space isomorphisms and operators}
\labl{sc:HilbertIsos}

To facilitate the computation of traces over orbifold Hilbert spaces 
later, it is useful to identify isomorphisms between some relevant
Hilbert spaces. On the level of the torus $\Tors$ we have the
isomorphism 
\equ{
\cT:~ \cH_{\Tors} \ra \cH_{\Tors,T}; 
\qquad 
(\cT \gps)(x) = \cT(x) \gps(x). 
\labl{IsoTors}
}
Here the function $\cT(x)$ is obtained as follows: Since all $T_v$
commute \eqref{consistency} for all $v \in \gG$, it follows that there
are commuting generators $H_I$ such that 
$T_v= \exp({i\, a^T \get\, v})$ where 
$a^\gm = \sum_I a^{\gm\, I} H_I$ and $\get$ represent the Minkowskian
metric. Using this, we can easily construct a function $\cT(x)$ which
is $\gG$--periodic up to $T$ transformations: 
\equ{
\cT(x) = e^{i\, a^T \get \, x},
\qquad 
\cT(0) = 1, 
\qquad 
\forall v \in \gG:~ 
\cT(x+v) = T_v\, \cT(x).
\labl{SolvePeriodicity}
}
In other words, the function $\cT(x)$ can be used to turn periodic
functions into functions periodic up to $T_v$ transformations, and
vice versa. The projection equation \eqref{Projgv} at
$\Fixd_{\,g}$ can be represented as 
\equ{
\cR_g(x) \, \gps(x) = \gps(x), 
~ \forall x \in \Fixd_{\,g}; 
\quad \text{with} \quad 
\cR_g(x) =  \cT{}^\perp_{\,g}(x)\, \cT{}^\perp_{\,g}(g\, x)\inv\,  R_g.
\labl{ProjgvGen} 
}
To see that this is equivalent with the condition given above, notice
that for $x \in \Fixd_{g\,s}$: 
\(
\cR_g(x) = \cR_g(\fZ{}_g^s) = 
 \cT{}^\perp_g(\fZ{}_g^s)\, \cT{}^\perp_{\,g}(g\, \fZ{}^s_g)\inv\,  R_g
= R{}^s_g.
\)
This notation can be used to avoid having to specify with which
subspace $\Fixd_{g\,s}$ of $\Fixd_{\,g}$ one is concerned with.

In \eqref{InnerOrbi} the inner product on the orbifold Hilbert space
$\cH_{\Tors/G,T,R}$ had been expressed in terms of the inner product
on the torus Hilbert space. However, not all states in the torus
Hilbert space $\cH_{\Tors,T,R}$ descent down to states on the
orbifold. To obtain the states which do, we define the orbifold
projection operator 
\equ{
\cP_R :~ \cH_{\Tors,T} \ra \cH_{\Tors,T}; 
\qquad 
(\cP_R \gps)(x) = 
\frac 1{|G|} \sum_{g \in G} R_g \, \gps(g\inv x) 
\labl{OrbiProj}. 
}
For $\gps \in \cH_{\Tors,T}$, i.e.\ $\gps(x+v) = T_v \gps(x)$ it
follows indeed that 
\equ{
(\cP_R \gps)(h\, x) = R_h \, (\cP_R \gps)(x), 
\qquad 
(\cP_R \gps)(x + v) = T_v\,  (\cP_R \gps)(x), 
\labl{OrbiTwistStates}
}
because any $h \in G$ defines a group automorphism by $h(g) = hg$,
$R$ is a homomorphism \eqref{homom}, and formula
\eqref{consistency} for $T_{g\inv v}$. Moreover, the operator $\cP_R$
satisfies the usual properties of a projection operator 
\equ{
\cP_R{}^2 = \cP_R{}^\dag = \cP_R. 
\labl{ProjProps}
}
These properties follow, using again that any $h \in G$ defines an
automorphism on $G$. In addition, the Hermiticity property required  a
change of variables in the integral over $\Tors$ within the Hilbert
space inner product. The natural isomorphism 
$\cH_{\Tors,T,R} \ra \cH_{\Tors/G,T,R}$ allows us to perform all
calculations in the orbifold Hilbert space in the Hilbert space of 
torus states with appropriate boundary conditions instead. In fact, 
we may simply work with the states of the Hilbert space
$\cH_{\Tors,T}$, and use the operator $\cP_R$ to project on to
$\cH_{\Tors,T,R}$. Since the former has a much simpler structure, this
proves to be a very convenient strategy.

Finally, we discuss some properties of general operators on
$\cH_{\Tors,T}$ and $\cH_{\Tors/G,T,R}$, which turn out to be of central
importance in the subsequent discussion. Let $\cO$ be an operator on
$\cH_{\Tors,T}$. Its expectation value in the state $|\gps \rangle$ is
denoted by  
\(
\langle \gps \,|\, \cO \,|\, \gps \rangle_{\Tors,T}. 
\)
A coordinate space representation of this operator is written as  
$\cO(x, \der_x)$. To avoid that the derivatives act on the operator
itself, we assume that the operators $x$ and $\der_x$ are ordered such
that the operators $\der_x$ stand on the right: 
\equ{
\cO(x, \der_x) = \sum_{\vec{\gm}} \cO^{\vec{\gm}}(x)\, 
\der_{\vec{\gm}} = 
\sum_k \sum_{\gm_1\ldots \gm_k} 
\cO^{\gm_1\ldots \gm_k}(x)\, \der_{\gm_1}\ldots \der_{\gm_k}, 
\labl{DefOper} 
}
where the sum is over all multi--indices 
$\vec{\gm}= (\gm_1, \dots, \gm_k)$ and 
$\der_{\vec{\gm}} = \der_{\gm_1}\ldots \der_{\gm_k}$ for all 
integral $k \geq 0$, and $\gm_i = 0,\ldots, d-1$. In general the
different $\cO^{\vec{\gm}}(x)$ do not mutually commute, 
and may also not commute with $R_g$ and $T_v$.

Since the orbifold Hilbert space can be viewed as a subset of the
torus Hilbert space, also not all operators on the torus Hilbert space
make sense on the orbifold Hilbert space. Those operators that do, are
called compatible with the orbifold boundary conditions. To
investigate what the defining property of such operators is, let 
$\cO:~ \cH_{\Tors,T} \ra \cH_{\Tors,T}$ and $\gps \in \cH_{\Tors,T,R}$
be an orbifold state, i.e.\ $\gps(g\, x) = R_g\, \gps(x)$. If $\cO$ is
an orbifold compatible operator, then also $\cO\, \gps$ should be an
element of $\cH_{\Tors,T,R}$. By applying the orbifold boundary
conditions to $\cO\, \gps$, we find that  
\equ{
\cO(g\,x, \der_x\, g\inv) = R_g\, \cO(x, \der_x)\, R_g{}\inv. 
\labl{OrbiOperator}
}
From any operator $\cO:~\cH_{\Tors,T} \ra \cH_{\Tors,T}$ an orbifold
compatible operator $\cO_R$ can be obtained by 
\equ{
\cO_R(x, \der_x) = \frac 1{|G|} \sum_{h\in G} 
R_h{}\inv \cO(h\,x, \der_x\, h\inv)\, R_h. 
\labl{MakeOrbiOperator}
}
In fact, this operator will arise naturally when we compute traces on
orbifolds, see section \ref{sc:main}. If $\cO$ is an orbifold
compatible operator, it follows trivially that $\cO_R = \cO$.

\section{Orbifold Hilbert space traces and their applications}
\labl{sc:main}

In this section we expose our main result concerning the traces on
orbifolds: It describes how the trace of an operator on a Hilbert
space of a non--trivial orbifold bundle, can be reduced to a sum over
traces over Hilbert spaces associated to trivial ones. The
various technicalities of the computations are collected in appendix
\ref{sc:traces}.  After this general exposition of the computation of
Hilbert space traces on orbifolds, we discuss a few possible
applications of this general method: In subsections \ref{sc:Anomalies}
and \ref{sc:Tadpoles} we consider anomalies, and tadpoles on
orbifolds, respectively.

\subsection{Tracing over the orbifold Hilbert space}
\labl{sc:OrbiTrace}

A orthonormal basis $\{ | \gf_\gs(q) \rangle\}$ for the Hilbert space
$\cH_\Tors$ has been constructed in appendix \ref{sc:scalar}, see eq.\
\eqref{BasisHilbert}. Here $\gs$ labels the basis vectors $\ge_\gs$ of
the fiber $\Fibr$, and $q \in \Tors^*$ takes its values in the vector
space dual to $\Tors$. It should be noted that none of our results
depend on this explicit basis. 
We define the trace $\Tr_{\Tors}$ of an operator $\cO$ on the Hilbert
space $\cH_\Tors$ associated to a trivial bundle over a torus $\Tors$
as    
\equ{
\Tr_{\Tors}\,[ \cO ] =  
\sum_\gs \int\limits_{\Tors^*} \d q\, 
\langle \gf_\gs(q) \,|\, \cO \,|\, \gf_\gs(q) \rangle_{\Tors}
= 
\sum_\gs 
\int\limits_{\Tors\otimes \Tors^{*}} \!\! \d x \, \d q~
\gf_\gs(x;q)^\dag \, \cO(x, \der_x) \, \gf_\gs(x;q).
\labl{TrTorsTrace}
}
For fermionic fields there will be an additional minus sign; we will
always write it explicitly. 
In general such traces are defined only formally, as they may not
converge and regularization is required. In the following we assume
that either the operator $\cO$ is a bounded operator, or that it
has been regularized such that it has become a bounded operator. 
From this trace on a trivial bundle over the torus, we can
define the trace on a torus bundle with periodicity conditions
\eqref{TorsBound} defined by the homomorphism $T$. Because of the
isomorphism \eqref{IsoTors} between  $\cH_\Tors$ and $\cH_{\Tors,T}$,  
the trace on $\cH_{\Tors,T}$ reads: 
\equ{
\Tr_{\Tors, T}\,[ \cO ] = \Tr_{\Tors}\,\Bigl[ \cT^\dag\, \cO \, \cT \Bigr].
}
In turn, we can use this to define the trace $\Tr_{\Tors/G,T,R}$
on the non--trivial orbifold bundle defined by the orbifold twist
$R$, exploiting the defining isomorphism between $\cH_{\Tors/G,T,R}$
and $\cH_{\Tors,T,R}$. Let $\cO$ be an compatible operator
\eqref{OrbiOperator} on $\cH_{\Tors,T,R}$, so that it can be
restricted to the orbifold. We can compute the trace of $\cO$ on 
$\cH_{\Tors/G,T,R}$ as 
\equ{
\Tr_{\Tors/G,T, R}\,[ \cO ] = \frac 1{|G|}\Tr_{\Tors,T, R}\,[ \cO ] = 
\frac 1{|G|} \Tr_{\Tors,T}\,\Bigl[ \cP_R{}^\dag \, \cO \, \cP_R \Bigr] = 
\frac 1{|G|} \Tr_{\Tors}\,\Bigl[ 
\cT^\dag \cP_R{}^\dag \, \cO \, \cP_R \cT
\Bigr],
\labl{OrbiTraceDef}
}
with the help of the orbifold projector $\cP_R$, defined in 
\eqref{OrbiProj}. This trace can be evaluated by using the
definition of the traces of the trivial bundle on the torus
\eqref{TrTorsTrace} and by substituting the definition of the orbifold
projector \eqref{OrbiProj}.

The details of this calculation are rather lengthy, and are therefore
given in appendix \ref{sc:traces}; we only give the main results
here. We also give some intermediate results, which are slightly more
general than required by the calculation at hand: they often apply to
any operator on $\cH_{\Tors,T}$, not only the orbifold compatible
ones. First of all, for any operator on $\cH_{\Tors,T}$ the
relation    
\equ{
\Tr_{\Tors,T, R}\,[ \cO] =  
\Tr_{\Tors, T}\,\Bigl[\cP_R{}^\dag \, \cO_R\Bigr]
\labl{TraceOrbiComp}
}
holds, where $\cO_R$ is the orbifold compatible operator defined in 
\eqref{MakeOrbiOperator}. Also for any operator on $\cH_{\Tors, T}$ we
find that 
\equ{
\Tr_{\Tors,T}\,\Bigl[\cP_R{}^\dag \cO \Bigr] = 
\frac 1{|G|} \sum_{g \in G} \frac 1{|\det{}^\perp_{\,g}(1-g)|} 
\Tr_{\Fixd_{\,g},T{}^\parl_g}\, \Bigl[ \cR_g \, \cO_g \Bigr], 
\labl{RemoveOrbiProj}
}
using the definition \eqref{ProjgvGen} and a rescaling of the
derivative perpendicular to $g$:
\equ{
\cO_g(x, \der) = 
\cO \bigl(x, 
\der{}_{g}^\parl + (\der (1- g)\inv ){}_{\,g}^\perp
\bigr).
\labl{Operh}
}
Here the same ordering as in \eqref{DefOper} has been applied. Notice
that for the identity: $\cO_{1}(x, \der_x) = \cO(x, \der_x)$. In
addition, we have used the notation defined in \eqref{IntFixdDelta} and
\eqref{SolvePeriodicity}. It should be noted that this result depends
on careful treatment of the properties of the delta function, see the
discussion leading to \eqref{InterRes} of appendix \ref{sc:traces}.

The result of the computation of \eqref{OrbiTraceDef} can be obtained
by combining  \eqref{TraceOrbiComp} and
\eqref{RemoveOrbiProj} for an orbifold compatible operator $\cO$ on
$\cH_{\Tors,T}$, for which $\cO_R = \cO$ see discussion below
\eqref{MakeOrbiOperator}. Making use of the discussion of the
orbifolding of fixed spaces at the end of subsection \ref{sc:geom}, we
can write the trace over the orbifold Hilbert space as  
\equ{
\Tr_{\Tors/G,T, R}\,[ \cO ] = 
\frac 1{|G|} \sum_{g_r} 
\frac 1{|\det^\perp_{g_r} (1\!\!-\!g_r)|} 
\sum_{s_r} 
\Tr_{\Fixd_{g_r\,s_r}/H_{g_r}^{s_r}, T{}^\parl_{g_r}}\, 
\Bigl[ R_{g_r}^{s_r}\,  \cO_{g_r} \Bigr],
\labl{OrbiTrResult}
}
with $R^{s_r}_{g_r} = {T_{v{}_{g_r}^{s_r}}}\inv R_{g_r}$. 
The sums here are over the representatives $g_r$ of the conjugacy
classes $(g_r)$ of the group $G$, and the representatives $s_r$ of the
equivalence classes of the fixed point labels. 
Notice that if the homomorphism $T$ is the identity, we would find
$R_{g_r}^{s_r} = R_{g_r}$ at all fixed spaces 
$\Fixd_{g_r\,s_r}/H_{g_r}^{s_r}$. 
This shows, that the non--periodic torus boundary conditions, encoded by
the non--trivial homomorphism $T$, lead to modifications of the traces at
fixed spaces compared to those trace on a pure orbifold.

The remaining traces are rather straightforward since they are over
states which are periodic on $\Fixd_{g_r\,s_r}$ up to transformations 
$T{}^\parl_g$ only.  However, there is no orbifold projection in the
fiber of the bundle anymore: The orbifold twist is encoded entirely in
the matrix  $R{}_{g_r}^{s_r}$. To work these traces out further we
need more information concerning the structure of the relevant
fibers. However, since at fixed point $s_r$ the matrix
$R{}_{g_r}^{s_r}$ generates an Abelian group of order $|g_r|$, we may
decompose this matrix  
\equ{
R{}^{s_r}_{g_r} = \sum_{k=0}^{|g|-1} 
e^{2\pi i\, k/|g|} \,  P{}^{s_r\, k}_{g_r}
\labl{Projdecomp}
}
in terms of the projection operators $P{}^{s_r\, k}_{g_r}$ of the
possible $|g|$, eigenvalues $\exp(2\pi i\, k/|g|)$, of $R{}^{s_r}_{g_r}$.

\subsection{Anomalies}
\labl{sc:Anomalies}

As mentioned in the introduction of this paper, there has been a lot 
of attention to the calculation of anomalies on different types of 
orbifolds. We will not compare our general results with all of them, 
but rather we focus on the following question: It is well--know that 
only in even dimensions there are gauge (and gravitational) 
anomalies. On the other hand it has been argued by various authors 
that on the orbifold $S^1/\Intr_2$ gauge anomalies arise at the fixed 
points. \footnote{By Arkani--Hamed et al.\ \cite{Arkani-Hamed:2001is}
the axial anomaly has been computed from first principles; but not the
gauge anomaly. Even though both anomalies are similar in structure,
there are also  important differences.} Only very recently a more general
analysis of anomalies on orbifolds has been given in ref.\
\cite{vonGersdorff:2003dt}.  Here we would like to present a description 
of gauge anomalies that applies to any orbifold of the type 
discussed in subsection \ref{sc:geom}. In particular, we allow that 
the full space and the fixed hyper surfaces can all have either 
even or odd dimensionality.

We consider a fermion $\gps$ on $\Tors/G$ coupled to a
Hermitian (non--)Abelian gauge connection $A_\gm$ and a spin--connection
$\go_\gm$, as this is the typical situation in which anomalies may 
arise. In general, this fermion may transform under some non--Abelian 
(and chiral for even dimension $d$) gauge transformation, to 
which the gauge field $A_\gm$ is associated. On the $d$ dimensional
space $\Tors$ with Minkowskian signature, its (kinetic) action takes
the standard form   
\equ{
S = - \int \d^{d}x\, \bgps D\Slashed \gps, 
\qquad 
D\Slashed = \der\slashed + i A\Slashed + \go\Slashed. 
}
Of course the boundary conditions \eqref{BoundConds} only make sense, if
$T_v$ and $R_g$ define symmetries of the action of the fermion. Therefore
for all Minkowski indices $\gm = 0,1,\ldots d-1$ we have that 
\equ{
(-\gg_0 {T_v}^\dag \gg_0) \, \gg^\gm\, T_v = \gg^\gm, 
\qquad 
(-\gg_0 {R_g}^\dag \gg_0) \, {(g\inv)^\gm}_\gn\, \gg^\gn\, R_g = \gg^\gm.  
}
The gauge and spin--connections contained in the Dirac operator have
to satisfy corresponding boundary conditions for consistency; they
read
\equ{
\arry{c}{
(-\gg_0 {T_v}^\dag \gg_0) \, \gg^\gm\, A_\gm(x+v)\,  T_v = \gg^\gm \, A_\gm(x), 
\quad
(-\gg_0 {R_g}^\dag \gg_0) \,  {(g\inv)^\gm}_\gn\,\gg^\gn\, A_\gm(gx)\, R_g =
\gg^\gm\, A_\gm(x),   
}
\non \\ 
\arry{c}{
(-\gg_0 {T_v}^\dag \gg_0) \,\gg^\gm\, \go_\gm(x+v)\,  T_v = \gg^\gm \, \go_\gm(x), 
\\[2ex]
(-\gg_0 {R_g}^\dag \gg_0) \,{(g\inv)^\gm}_\gn\, \gg^\gn\, \go_\gm(gx)\, R_g = \gg^\gm\, 
\go_\gm(x).
}
\labl{GaugeComp}
}
The boundary conditions \eqref{BoundConds} may lead to chiral
fermions in complex representations at some of the even dimensional
fixed hyper surfaces. The conditions \eqref{GaugeComp} ensure that the
Dirac operator $D\Slashed$ is an orbifold compatible operator
\eqref{OrbiOperator}. It should also be realized that the transformation
parameter, $\gL^\dag = \gL$, satisfies corresponding boundary
conditions to the ones given in \eqref{GaugeComp} for the gauge and
spin--connections. In addition, there may be chiral and/or Majorana
projections that act on the spinors, depending on the 
dimensionality $d$.

Before continuing with the discussion of the anomalies on the
orbifold, let us mention some well--known results concerning anomalies
on a smooth manifold $\Mnfd$ of dimension $d$, i.e.\ without boundaries
and singular hyper surfaces. A constant parameter $\gL$ can be any
element of the $d$ dimensional Clifford algebra. However, since we
only consider gauge fields and the spin--connection, their parameters
are proportional to the identity and the spin generators 
$\frac 14 [ \gg_\gm, \gg_\gn ]$, 
respectively. Moreover, if the dimension $d$ of $\Tors$ is even, we
may allow that the (gauge) transformations are different on the left--
and right--handed fermionic components. 
Such transformations can be anomalous because the fermionic path
integral measure is not necessarily invariant under the fermionic
field redefinition $\gps \ra ( 1 + \gL) \gps$ and 
$\bgps \ra \bgps (1 -\gg_0 \gL^\dag \gg_0)$. This may lead to the
non--conservation of the gauge current $J^\gm$ on the manifold 
$\Mnfd$:
\equ{
\langle \der_\gm J^\gm \rangle_{\Mnfd} = 
\cA_{\Mnfd}(\gL) = 
- \Tr_{\Mnfd}\, 
\Bigl[ \,\gL \, e^{D\slashed^2/M^2} \Bigr] 
-
\Tr_{\Mnfd}\, 
\Bigl[\, -\gg_0 \gL \gg_0 \, e^{D\slashed^2/M^2} \Bigr],
\labl{DefAnom}
}
the minus signs arise because the trace is over fermionic degrees of
freedom. Here we have introduced the so--called heat kernel
regularization following the standard method of Fujikawa
\cite{Fujikawa:1980eg,Fujikawa:1979ay,Fujikawa:1984bg}; without any
regularization the traces of the gauge parameters $\gL$ and $-\gg_0 
\gL^\dag \gg_0$ over the full Hilbert spaces of $\gps$ and $\bgps$ are
ill--defined. The regulator parameter $M$ is taken to infinity at the
end of the anomaly calculation. The result is that only in even
dimensions there can be anomalies, and they are generally given by 
\equ{
\cA_\Mnfd(\gL) = 2\gp\, \int_\Mnfd \gO_{2k}^1(\gL; A, F, \go, R) 
}
where $F$ and $R$ are the field strengths of $A$ and $\go$,
respectively. The $d = 2k$ form $\gO_{2k}^1$ follows from the solution
of the descent equations, that solve the Wess--Zumino consistency
conditions \cite{Wess:1971yu},  
from the close and invariant $2k+2$ form 
$\gO_{2k+2}$ \cite{Zumino:1984rz,Zumino:1985ws}: 
\equ{
\gd_\gL \gO_{2k+2}(F,R) = \d \gO_{2k+2}(F,R) = 0, 
\\[1ex]
\d \gO_{2k+1}(A,F,\go,R) = \gO_{2k+2}(F,R), 
\qquad 
\gd_\gL \gO_{2k+1}(A,F,\go,R) = \d \gO_{2k}^1(\gL; A, F, \go, R).
\non
}
The precise particle content determines what the explicit form of the
defining polynomial $\gO_{2k+2}$ is; for a spin 3/2 or spin 1/2 field
in representation $\rep{R}$ of the gauge group we find  
\equ{
\gO^{3/2}_{2k+2} = \left. \hat A_{3/2}(R) \right|_{2k+2}, 
\qquad 
\gO^{1/2,\rep{R}}_{2k +2} = 
\left. \hat A_{1/2}(R) \tr_{\rep{R}} e^{i F/2\gp}\right|_{2k+2},
}
respectively. Here $\hat A_{s}$ with $s=1/2,3/2$ refer to the roof
genus, for their definitions see 
\cite{Alvarez:1984yi,Alvarez-Gaume:1984ig,gsw_2,Nakahara:1990th}.

Let us now return to the orbifold. Because both the gauge parameter
$\gL$ as well as the Dirac operator $D\Slashed$ are compatible with
the orbifolding, the traces in the definition of the anomaly
\eqref{DefAnom} are well--defined. Therefore we may apply
\eqref{OrbiTrResult} to obtain 
\equ{
\cA_{\Tors/G,T, R}  = 
\frac {-1}{|G|} \sum_{g_r, s_r} 
\frac 1{|\det^\perp_{\,g_r} (1\!\!-\!g_r)|} 
\Tr_{\Fixd_{g_r\, s_r}/H^{s_r}_{g_r}, T{}^\parl_{g_r}}
\!\!
\left[  
\Bigl(  
R{}^{s_r}_{g_r} \gL \!-\! \gg_0 R{}^{s_r}_{g_r}{}^{\!\dag}  \gL^\dag \gg_0 
\Bigr) 
e^{D\slashed^2/M^2} 
\right]_{g_r}\!\!.
\labl{OrbiTrAnom}
}
A comment is in order here: In this calculation we have taken $\gps$
and $\bgps$ to be independent. Therefore, we not only have to take
into account how the orbifold operator $\cR_g$ acts on $\gps$, but we
also remember that on $\bgps$ it acts as 
$- \gg_0 R{}^{s_r}_{g_r}{}^{\!\dag} \gg_0$.

To express this in terms of the anomaly forms $\gO_{2k}$, we need in
general more information concerning the precise structure of the
bundle in which the fermions take their value. The method of
decomposing the $R^{s_r}_{g_r}$ explained in \eqref{Projdecomp} can
also be applied here.

\subsection{Tadpoles and vacuum energy}
\labl{sc:Tadpoles}

As a second example of the possible one--loop quantities one can
compute on orbifolds using the general machinery exposed in this
paper, we would like to mention the computation of one--loop tadpoles
and the vacuum energy. These quantities can often be very significant
to understand the physics of the underlying theory. In particular,
possible (local) Fayet--Iliopoulos tadpoles in supersymmetric theories
in higher dimensions may trigger spontaneous symmetry breaking. 
This may also lead to localization of charged zero modes
\cite{GrootNibbelink:2002qp}. These types of tadpoles in general can
introduce quadratic sensitivity to a high scale in supersymmetric
theories of extra dimensions \cite{Ghilencea:2001bw,Barbieri:2001cz}. 
Here, we do not aim at full generality, since the computation of both
these quantities on regular spaces are well--known; rather we would
like to illustrate how such quantities can be computed on
general orbifolds. The methods described here can be used to confirm
the recent results obtained in ref.\ \cite{GrootNibbelink:2003gb}, in
which tadpoles on the orbifold $T^6/\Intr_3$ were computed.

Consider a theory with $N$ real scalars $\gf^\ga$ with a potential 
$V(\gf)$. Let us decompose the scalars in their background values
$\gf_0^\ga$ and their quantum fluctuations $\gd \gf^\ga$. The
vacuum energy $V(\gf_0)$ and the one--loop tadpoles $T_\ga(\gf_0)$ 
read
\equ{
V(\gf_0) = \frac 12 \Tr\, \ln 
\Bigl[ - \Box + M^2(\gf_0) \Bigr],
\quad  
T_\ga(\gf_0) = 
\frac 12\, \Tr \Bigl[
(M^2)_{,\ga}(\gf_0) \frac 1{- \Box + M^2(\gf_0)}
\Bigr], 
}
with the Euclidean Laplacian $\Box = \der_\gm^2$ and 
the mass matrix $(M^2)_{\ga\gb}(\gf_0) = V_{,\ga\gb}(\gf_0)$.
As explained in ref.\  \cite{Sher:1989mj}, and references therein, the
effective potential can be obtained from the tadpoles by integration
over the background fields. The definitions above are ill--defined and
require regularization. Various methods, like dimensional,
zeta--function, or heat--kernel regularization can be applied to do
this; see for example ref.\ \cite{Ghilencea:2002ff} for a discussion
how these different schemes are related. Here we restrict ourselves to
the zeta--function regularization method, and define 
\equ{
I_\gz(\ga, M^2; \gd) = 
\Tr \Bigl( - \Box + M^2 \Bigr)^{-\ga-\gd},
\labl{zeta}
}
with $\ga$ an arbitrary real constant and the regulator 
$\gd \in \Cplx$, which should be taken to zero at the end of a
physical calculation. The regulated expressions for the tadpoles and
the vacuum energy read 
\equ{
T_\ga(\gf_0; \gd) = - \frac 1{2\gd} \frac{\gd}{\gd \gf_0^\ga}\,
I_\gz(0, M^2(\gf_0); \gd), 
\qquad 
V(\gf_0; \gd) = - \frac 1{2\gd} \,  I_\gz(0, M^2(\gf_0); \gd). 
}
These quantities can now be computed on the orbifold using eq.\
\eqref{OrbiTrResult}, and expressed as a sum of traces on the
different spaces $\Fixd_{g_r\,s_r}/H_{g_r}^{s_r}$.

\section{Examples}
\labl{sc:Examples}

In this section we give three examples of orbifolds to which the
general procedure for computing orbifold traces, developed in the
preceding sections, can be applied. The first example we
discuss is the well--known orbifold $S^1/\Intr_2$. As this is a very
simple and well--studied orbifold, our general methods actually
provide a bit of overkill in this case. However, because of its 
familiarity it might help the reader to understand the general
features of the procedure advocated in this work. The next example
concerns a ten dimensional model, for example a super Yang--Mills
theory or supergravity, on the orbifold $T^6/Z_4$. The third
example we consider, is the non--Abelian orbifold $T^4/D_4$ with the
$D_4$ the dihedral group with eight elements. 

\subsection{5D theories on the orbifold $\boldsymbol{S^1/\Intr_2}$}
\labl{sc:S1Z2}

As stated in the introduction of this section, this subsection is
meant to illustrate the general machinery presented in the
preceding part of this article. Consider five dimensional Minkowski
space $\Real^{1,4}$, the one dimensional lattice and the group  
\equ{
\gG = \pmtrx{0_4 \\  \Intr} 
\qquad 
G = \{ \Id_5, g \} \cong \Intr_2, 
\qquad 
g = \pmtrx{\Id_4 & \\ & \mbox{-}1},
}
where $S^1 = \Real/\Intr$. 
As the group $G$ is Abelian, it follows that the group elements form
the conjugacy classes, and the centralizer is the group itself. The
only non--trivial projection operators on 
$\Tors = \Real^{1,3}\times S^1$ read 
\equ{
P{}^\parl_g = \pmtrx{ \Id_4 & \\ & 0 }, 
\quad 
P{}^\perp_{\,g} = \pmtrx{ 0_4 & \\ & 1 }
~ \Ra ~ 
\Tors{}^\parl_g = \pmtrx{\Real^{1,3} \\ 0}, 
\quad 
\Tors{}^\perp_{\,g} = \pmtrx{0_4 \\ S^1}, 
\qquad s = 0,1. 
}
In particular we find that $\det{}^\perp_g(1-g) = 2$. 
The fixed points and fixed sets are easily identified 
\equ{
\fZ_g^s = \pmtrx{ 0_4 \\ \frac 12\, s }, 
\qquad 
v{}^s_g = \pmtrx{ 0_4 \\ -s }, 
\qquad 
\Fixd_{g\, s} = \fZ_g^s + \pmtrx{ \Real^{1,3} \\ 0 }.
}
The mapping $p_g^\parl$ defined in \eqref{Residual} is trivial 
\equ{
p_g^\parl(g) = P{}^\parl_g, 
\qquad 
\text{Ker}^\parl_g = G.
}
For the fixed sets in the orbifold we thus find the following 
\equ{
\Fixd_{\, 1}/G = \Tors/G = \Real^{1,3} \times S^1/\Intr_2,
\qquad 
\Fixd_{\, g}/G = \sum_{s=0,1} \Fixd_{g\, s}.   
}
This concludes the geometrical description of the orbifold.

As we described in subsection \ref{sc:FiberInner} any field theory on
this orbifold is essentially determined by two group homomorphisms $T$
and $R$ of $\gG$ and $G$, respectively, to the group of
diffeomorphisms of the relevant fiber vector space in which the field
takes its values: 
\equ{
\psi(x + 2\pi) = T\, \psi(x), 
\quad
\psi(- x) = R\, \psi(x),
\qquad 
R^2 = (TR)^2=1. 
}
In the general trace formula on this orbifold this can be left
unspecified, and we obtain:  
\equ{
\Tr_{\Tors/G, T, R}\Bigl[ \cO \Bigr] = 
\frac 12 \Tr_{\Tors/G, T}\Bigl[ \cO \Bigr] + 
\frac 12 \cdot \frac 12 \sum_{s=0,1} 
\Tr_{\Fixd_{g\, s}} \Bigl[ R{}^s_g \cO_g \Bigr],
\labl{S1Z2TrResult}
}
where $\cO_g(x^i, x^5; \der_i, \frac 12\, \der_5)$ and 
$R_g^s = T^{-s} R = R T^s$.

Let us work out this formula for the example of the gauge anomaly
of a single fermion $\gps$ 
on this orbifold coupled to an Abelian gauge field $A_\gm$ and gauge
parameter $\gL$. Upon requiring that there is a four dimensional gauge
field zero mode, the orbifold twist action on these fields is uniquely
determined to be  
\equ{
\arry{c}{
\gps(x +2\pi) = (-)^p \gps(x), 
\quad 
\bgps(x +2\pi) = (-)^p \bgps(x), 
\\[1ex]  
A_i( x+2\pi) = A_i(x), 
\quad 
A_5(x+2\pi) = A_5(x), 
\\[1ex]
\gps(g\, x) = \gg_5\, \gps(x), 
\quad 
\bgps(g\, x) = \bgps(x) (- \gg_5), 
\quad 
A_i(g\, x) = A_i(x), 
\quad 
A_5(g\, x) = - A_5(x),
}
} 
here $p =0,1$ specifies whether we take periodic ($p=0$) 
or anti--periodic ($p=1$) 
boundary conditions on the circle. (For a single fermion $T$ and $R$
necessarily commute, hence $T^2=1$. Requiring that the fermionic
action is invariant then implies that $T = (-)^p$.)
In the second relation we have used
that  $-\gg_0 R_g{}^\dag \gg_0 = - \gg_0 \gg_5 \gg_0 = - \gg_5$. 
In equation \eqref{OrbiTrAnom} for the general anomaly the part in
between the round brackets becomes for the only non--trivial element
$g$ 
\equ{
 \cR_{g}^s i \gL + \gg_0 \cR_g^s{}^\dag i \gL \gg_0 = 
2 (-)^{ps} \gg_5 i \gL.
}
As there are no anomalies on odd dimensional spaces when the spin
bundle is trivial, it follows that only the trace over the four
dimensional part can give rise to an anomaly. Confirming the by now
well--know results:
\equ{
\cA[\gL] = \frac 12 \sum_{s=0,1} 2\pi 
\int\limits_{\Real^{1,3}\otimes S^1/\Intr_2} 
(-)^{ps}\, \gO_{4}^1(\gL; A, F) \, \gd( x^\perp_{\, g} - s \,\gp).
}
Here we have used that the standard gauge anomaly in four dimensions
is defined w.r.t.\ $(1+\gg_5)/2$ left--handed chirality states, while
here we find the trace over $\gg_5$. This cancels one factor of $1/2$
in the general trace formula \eqref{S1Z2TrResult} on this orbifold.
For the case of anti--periodic fermions on the circle, we see that 
the anomalies at both fixed points are opposite.

\subsection{10D theories on the orbifold $\boldsymbol{T^6/\Intr_4}$}
\labl{sc:T6Z4}

Next we start with ten dimensional Minkowski space $\Real^{1,9}$ on
which we take partly complex coordinates: 
$(x, z_1, z_2, z_3) \in \Real^{1,3} \times \Cplx^3$. The lattice for
this example is taken to be $\gG = (0_4, \Intr^3 + i \Intr^3)$. And the
group $G \cong \Intr_4$ is generated by the action 
\equ{
g (x, z_1, z_2, z_3) = (x,- z_1, i\, z_2, i\, z_3), 
\qquad 
g^2 (x, z_1, z_2, z_3) = (x, z_1, - z_2, - z_3);
\labl{Z4twist}
}
the element $g^2$ generates a $\Intr_2$ subgroup of $G$. As in the
previous section the conjugacy classes are the elements of the group
themselves, since the group is Abelian. However, the fixed point
structure is more interesting in this case. First of all the
projectors defined in \eqref{gProj} read in this case 
\equ{
\arry{c}{
P{}_1^\parl (x, z_1, z_2, z_3) = (x,z_1, z_2, z_3),
\qquad 
P{}_g^\parl (x, z_1, z_2, z_3) = (x,0, 0, 0),
\\[2ex] 
P{}_{g^2}^\parl (x, z_1, z_2, z_3) = (x,z_1, 0, 0),
}
}
and $P^\parl_{g^3} = P^\parl_g$. Hence we see that the fixed
spaces of $g$ and $g^3$ are isomorphic to multiple copies of four
dimensional Minkowski spaces, while the fixed space $\Fixd_{g^2}$ is
six dimensional. To describe these fixed point spaces further, 
we introduce the following definition for the fixed points 
\equ{
\fZ_g^{p\, q} = 
(0, \gz_{p_1q_1}, \gz_{p_2p_2}, \gz_{p_3p_3}), 
\qquad 
\fZ_{g^2}^{p\, q} = 
(0, 0, \gz_{p_2q_2}, \gz_{p_3q_3}),
\qquad 
\gz_{p\,q} = \frac{p+i\, q}2, 
\labl{T6Z4Fixd}
}
with $p_j, q_j = 0, 1,$ of complex codimension three and two,
respectively. The fixed point sets take the form  
\equ{
\Fixd_{\, g} = \sum_{p,q} \Fixd_{g\, p,q}, 
\quad 
\Fixd_{g\, p,q} = \fZ_g^{p\, q} + \Tors^\parl_g, 
\qquad 
\Fixd_{g^2} = \sum_{p,q} \Fixd_{g^2\, p,q}, 
\quad 
\Fixd_{g^2\, p,q} = \fZ_{g^2}^{p\, q} + \Tors^\parl_{g^2}. 
}
Contrary to the previous example, now the equivalence relation of
fixed point labels is not entirely trivial: For $p\neq q$ we find
trivially that $ g\, \fZ^{p,q}_{g^2} = \fZ^{q, p}_{g^2}$ up to lattice
shifts; hence on the level of the full orbifold the corresponding
fixed point spaces are identified. As discussed in subsection
\ref{sc:geom} the precise prescription of this identification is
determined by \eqref{Residual}. The element $g$ acts non
trivially on the fixed spaces of $g^2$, in particular, we find that 
\equ{
p_{g^2}^\parl(g) (x, z_1) =  (x, - z_1). 
}
In words this means that the torus $T^2$ above $\fZ^{p,q}_{g^2}$ is
identified with the torus $T^2$ above $\fZ^{q, p}_{g^2}$ with opposite
orientations. Because of this identification, we define the sum over
the indices $p,q$ for the fixed point sets of $g^2$ up to the
interchange of these indices. In particular we will write $p\neq q$ to
indicate that the indices are not equal.

With this in mind let us turn to the orbifolding of the fixed point
sets. For the fixed points of $g$, and $g^2$ with $p\neq q$, 
the situation is again rather easy, since 
\equ{
G{}_g^{p,q} = \text{Ker}{}^\parl_{g\, p,q} \cong \Intr_4, 
\qquad
G{}_{g^2}^{p\neq q} = \text{Ker}{}^\parl_{g^2\, p\neq q} \cong \Intr_2. 
}
(Here we have used that $g$ does not map the fixed space of $g^2$ with
$p\neq q$ to itself.) Hence on the corresponding fixed points there is
no residual orbifolding. For the fixed points of $g^2$ with $p = q$
the situation is different since there
\equ{
G{}_{g^2}^{p= q} = \langle g \rangle \cong \Intr_4, 
\qquad 
\text{Ker}{}^\parl_{g^2\, p =q} = \langle g^2 \rangle \cong \Intr_2. 
}
This implies that 
$H{}_{g^2}^{p= q} \cong \Intr_4/\Intr_2 \cong \Intr_2$, and hence the
orbifold twist on those fixed points leads to the orbifold 
$T^2/\Intr_2$ above the fixed points $\fZ{}_{g^2}^{p= q}$. 
Before we can apply the general formula for the orbifold trace
\eqref{OrbiTrResult}, we need to determine the matrices 
$((1-h)\inv){}^{\perp}_h$ and the determinants 
$\det{}^{\perp}_{\,h}  (1-h)$ for $h = g, g^2$. 
Remembering that we use here a complex basis while 
$\det{}^\perp_{\, h}(1-h)$ is the real determinant, they read
\equ{
\arry{lcl}{
((1-g)\inv){}^{\perp}_{\,g} = \frac 12\, \text{diag}( 1, 1+ i, 1 + i), 
& \qquad &
((1-g^2)\inv){}^{\perp}_{g^2} = \frac 12\, \text{diag}(1, 1), 
\\[2ex]
\det{}^{\perp}_{\,g}  (1-g) = |2|^2 |1-i|^2  |1-i|^2  = 16, 
& \qquad & 
\det{}^{\perp}_{g^2}  (1-g) = |2|^2 |2|^2 = 16.  
}
}

On a given field $\gps$ the torus and orbifold boundary conditions read
\equ{
\gps(x + e_j) = T_{j}\, \gps(x),
\quad 
\gps(x + i\,e_j) = T_{j'}\, \gps(x),
\quad 
\gps(g\, x) = R\, \gps(x),
}
where for $j = 1,2,3$ the vectors $e_j$ denote the basis elements for
the complex three torus $T^6$. The consistency of the orbifolding
conditions with the periodicity conditions leads to requirements on
the matrices $T_j, T_{j'}, R$. For $j = 2,3$ they take the form 
\equ{
T_{j'} = R T_j R\inv, 
\qquad 
R^4= (T_j R)^4 = (R^2 T_{j})^2 = 1. 
}
This shows that the translations $T_{j'}$ and $T_j$ are not
independent. On the other hand, as the $g$ act as a $\Intr_2$ action
on $z_1$  the matrices $T_1$ and $T_{1'}$ are independent, and satisfy
\equ{
(R T_{1})^2 = 1, 
\qquad  
(R T_{1'})^2 = 1. 
} 

Taking all this information into account, we can write down the expressions
for the trace of an arbitrary operator $\cO$ which is compatible with
this orbifold. The result can be stated as 
\equ{
\Tr_{\Tors/\Intr_4, T, R}\, \Bigl[ \cO \Bigr] = ~
\frac 14 \Tr_{\Tors/\Intr_4} \Bigl[ \cO_1 \Bigr]
+ \frac 14 \cdot \frac 1{16} \sum_{p,q} 
\Tr_{\Fixd_{g\, p,q}}\Bigl[ R{}^{p,q}_{\, g} \cO_g \Bigr]
+ \frac 14 \cdot \frac 1{16} \cdot 
\\
\cdot 
 \sum_{p\neq q} 
\Tr_{\Fixd_{g^2\, p,q}}\Bigl[ R{}^{p,q}_{\, g^2} \cO_{g^2} \Bigr]
+ \frac 14 \cdot \frac 1{16}
\sum_{p=q} 
\Tr_{\Fixd_{g^2\, p,q}/\Intr_2}\Bigl[ R{}^{p,q}_{\, g^2} \cO_{g^2} \Bigr]
+ \frac 14 \cdot \frac 1{16} \sum_{p,q} 
\Tr_{\Fixd_{g^3\, p,q}}\Bigl[ R{}^{p,q}_{\, g^3} \cO_{g^3} \Bigr]. 
\non
}
Here the sums are only over the inequivalent fixed points. In this
formula we have repressed the explicit reference $T^\parl_g$. The 
operators $\cO_{g^i}$ are given by 
\equ{
\arry{c}{
\cO_{g} = 
\cO\Bigl(x,z;  \frac 12\der_1, \frac{1+i}2\der_2, \frac{1+i}2\der_3
\Bigr),
\quad 
\cO_{g^2} = 
\cO\Bigr(x,z; 
\der_1, \frac{1}2\der_2, \frac{1}2\der_3
\Bigr), 
\\[2ex]
\cO_{g^3} = 
\cO\Bigl(x,z; 
\frac 12\der_1, \frac{1-i}2\der_2, \frac{1-i}2\der_3
\Bigr), 
}
}
and of course $\cO_1 = \cO$. In writing the derivatives we have again
employed complex notation. Notice that if $\cO$ is an operator which
only depends on the Laplacian, $\cO_{g^3} = \cO_{g}$.

In this expression only the operators $R_g^{p,q}$, $R_{g^2}^{p,q}$ and
$R_{g^3}^{p,q}$ are left to be determined using their definitions 
\eqref{Projgv}. To this end we first calculate relevant shifts $v_g^s$
using \eqref{vgsDef} 
\equ{
\arry{lll}{
v_g^{p,q} &=  (0,-1,i,i) * \fZ_g^{p,q} - \fZ_g^{p,q} 
&=  (0, -p_1 - i q_1, -p_2, - p_3), 
\\[1ex]
v_{g^2}^{p,q} & = (0,0,-1,-1) * \fZ_{g^2}^{p,q} - \fZ_{g^2}^{p,q} 
& = (0, 0, -p_2 -i q_2, - p_3 - i q_3), 
\\[1ex]
v_{g^3}^{p,q} & = (0,-1,-i,-i) * \fZ_g^{p,q} - \fZ_g^{p,q} 
& = (0, -p_1 - i q_1, -i p_2, -i p_3). 
} 
}
Hence the matrices 
\equ{
\arry{llll}{
R^{p,q}_g &= T_{v^{p,q}_{g}}{}\inv R_g &= 
T_{(0, p_1 + i q_1, p_2, p_3)}\, R &= 
T_1^{p_1} \, T_{1'}^{q_1} \, T_2^{p_2} \, T_3^{p_3} \, R,
\\[1ex]
R^{p,q}_{g^2} &= T_{v^{p,q}_{g^2}}{}\inv R_{g^2} &= 
T_{(0, 0, p_2 + i q_2, p_3 + i q_3)}\, R &= 
T_2^{p_2} \, R T_{2}^{q_2} R\inv \,  
  T_3^{p_3}\,  R T_{3}^{q_3} R\inv \, R^2,
\\[1ex]
R^{p,q}_{g^3} &= T_{v^{p,q}_{g^3}}{}\inv R_{g^3} &= 
T_{(0, p_1 + i q_1, i p_2, i p_3)}\, R &= 
T_1^{p_1} \, T_{1'}^{q_1}\,  R T_2^{p_2} \,  T_3^{p_3} R\inv\, R^3,
}
}
encode all the possible effects of non--trivial periodicity boundary
conditions on $T^6$ that are compatible with the $\Intr_4$
orbifolding. (Of course, as soon as we take the field $\gps$ to be
periodic, i.e.\ $T_j = T_{j'}=1$, the traces at all fixed points within
a given class are identical: 
$ R^{p,q}_{g^k} = R^k$.)
In a future
publication \cite{Koba} we will apply these results to anomalies in
heterotic $\E{8}\times \E{8}'$ theory compactified on this orbifold
$T^6/\Intr_4$.

\subsection{The orbifold $\boldsymbol{T^4/D_4}$} 
\labl{sc:T4D4}

In our final example we consider a non--Abelian orbifold, with the
finite group $G$ isomorphic to the dihedral group $D_4$ with eight
elements. The space $\Tors = \Real^{1,7}/\gG$ is defined by the four
dimensional lattice, which is a complex basis takes the form 
$\gG = \Intr^2 + i\, \Intr^2$. On $\Tors$ the action of 
$G= \langle g, k \rangle \cong D_4$ is defined by the following
generating elements
\equ{
g = \pmtrx{ \Id_4 & & \\ & i & \\ & & \mbox{-}i }, 
\quad 
k = \pmtrx{ \Id_4 & & \\ & & 1 \\ & 1 & }, 
\qquad 
g^4 = k^2 = 1, 
\quad 
k g k\inv = g\inv. 
} 
Here we also gave their defining properties. 
The conjugacy classes $(h)$ of $h \in G$ are given by 
\equ{
\arry{c}{
(1) = \{ 1 \},
\quad  
(g) = \{ g, g^3 \}, 
\quad 
(g^2) = \{ g^2 \},
\quad 
(k) = \{ k, g^2 k \}, 
\quad 
(g k) = \{ g k, g^3 k\}. 
}
} 
As representatives of these conjugacy classes we take the elements in
between the round brackets. As in the previous two examples we need to
investigate the properties of these representatives and derived
objects. Since the analysis in this case is rather lengthy, we have
simply summarized the results in table \ref{tab:D4props}. 

\begin{table}
$
\renewcommand{\arraystretch}{.9}
\arry{l | c c c |  c c }{ 
&&&&& 
\\[-1ex]
h & 1 & g & g^2 & k & g k 
\\[1ex]\hline
& & & & & 
\\[-1ex] 
P{}^\parl_h 
& 
\pmtrx{ \Id_4 & & \\ & 1 & \\ & & 1}
& 
\renewcommand{\arraystretch}{0.9}
\pmtrx{ \Id_4 & & \\ & 0 & \\ & & 0}
& 
\renewcommand{\arraystretch}{0.9}
\pmtrx{ \Id_4 & & \\ & 0 & \\ & & 0}
& 
\frac 12 \pmtrx{ 2 \Id_4 & & \\ &  1& 1 \\ & 1 & 1}
& 
\frac 12 \pmtrx{ 2 \Id_4 & & \\ &  1& i \\ & \mbox{-}i & 1}
\\ & & & & & 
\\[-1ex] \hline 
&  & & & &
\\[-1ex]
d{}^\parl_h &
8 & 4 & 4 & 6 & 6 
\\[1ex] \hline 
&  & & & &
\\[-1ex]
( (1\mbox{-} h)\inv ){}^\perp_{\,h} &
- &
\frac 12 \pmtrx{ 1 + i & \\ & 1- i }
& 
\frac 12 \pmtrx{ 1 & \\ & 1 }
& 
\frac 12 \, P{}^\perp_{\, k} 
& 
\frac 12 \, P{}^\perp_{g k}
\\ & & & & & 
\\[-1ex] \hline 
&  & & & &
\\[-1ex]
\det^\perp_{\,h} (1 \mbox{-} h) & 
1 & 4 & 16 & 4 & 4
\\[1ex] \hline 
&  & & & & 
\\[-1ex]
C(h) & 
\langle g, k \rangle 
& 
\langle g \rangle 
& 
\langle g, k \rangle 
& 
\langle g^2, k \rangle 
& 
\langle g^2 \rangle 
\\[1ex] \hline 
&  & & & &
\\[-1ex] 
\text{Ker}{}^\parl_h & 
\{ 1 \} 
& 
\langle g \rangle 
& 
\langle g, k \rangle 
& 
\langle k \rangle 
& 
\{ 1 \} 
\\[1ex] \hline 
&  & & & &
\\[-1ex] 
\text{Fixed}
& 
\fZ_1 & \fZ_{g\, p} & \fZ_{g^2\, p,q} & \fZ_{k\, p,q} & \fZ_{gk\, p,q}
\\[-1ex] 
&  & & & &
\\[0ex] 
\text{points} & 0 
& (0_4, \gz_{p_1 p_1}, \gz_{p_2, p_2})
& (0_4, \gz_{p_1 q_1}, \gz_{p_2, q_2})
& \gz_{p\,q} ( 0_4, 1, \mbox{-}1 )
& \gz_{p\,q} ( 0_4, 1, i )
\\[1ex] \hline 
&  & & & &
\\[-1ex] 
\text{Shifts}\  v_{g}^s
& (0_4, 0, 0) 
& (0_4, \mbox{-} p_1, \mbox{-}i p_2) 
& \! (0_4, \mbox{-} p_1 \mbox{-} i q_1, \mbox{-} p_2 \mbox{-} i q_2) 
& \! (0_4, \mbox{-}p\mbox{-}iq, p\mbox{+}iq) 
& \!\! (0_4,  \mbox{-} p \mbox{-} i q, \mbox{-} i p \mbox{+} q) 
\\[-2ex] && &&&
}
$
\caption{
The properties of the representatives $h$ of the conjugacy
classes $(h)$ in $G$ have been collected in this table. The symbol 
$\gz_{p\, q}$ is defined as in eq.\ \eqref{T6Z4Fixd}.
}
\label{tab:D4props}
\end{table}

Not all the fixed points of $h$ given in table \ref{tab:D4props} are
independent, when we take the identification due to the centralizer
$C(h)$ of $h$ into account. The independent fixed points of the
various representatives of the conjugacy classes are 
\equ{
\fZ_{g\, p}: ~ \fZ_{g\, 00}, ~\fZ_{g\, 11}, ~\fZ_{g\, 01}; 
\quad 
\fZ_{g^2\, p q}: ~ \fZ_{00,00}, ~\fZ_{11,11}, ~\fZ_{00,11}, ~
\fZ_{01,01},~ \fZ_{01,10}, ~\fZ_{00,01},~ \fZ_{01,11}; 
\non \\[1ex]
\fZ_{k\, pq}: ~ \fZ_{k\,00}, ~\fZ_{k\,11}, ~\fZ_{k\,01}, ~\fZ_{k\,10}; 
\quad 
\fZ_{gk\, pq}: ~ \fZ_{gk\,00}, ~\fZ_{gk\,11}, ~\fZ_{gk\,01}, ~\fZ_{gk\,10}. 
\labl{indicesFixed} 
}

Let $\psi$ be a field on this orbifold satisfying the following
generating boundary conditions 
\equ{
\arry{ccc}{
\gps(x + e_j) = T_j\, \gps(x), 
&  \qquad &
\gps(x + i\, e_j) = T_{j'}\, \gps(x), 
\\[1ex] 
\gps(g\, x) = R\, \gps(x), 
& \qquad &  
\gps(k\, x) = S\, \gps(x),
} 
}
where $e_1, e_2$ denote the basis for the complex two torus $T^4$. 
The various consistency requirements of these transformations lead to 
constraints 
\equ{
R^4 = S^2 = 1, 
\quad 
S R S\inv = R\inv, 
\quad 
(R T_1)^4 = (R^2 T_1)^2 = 1
}
on $T_1$, $R$ and $S$. The other translation elements are not
independent, but are determined by 
\equ{
T_{1'} = R \, T_1\,  R\inv, 
\quad 
T_2 = S\, T_1 \, S\inv, 
\quad 
T_{2'} = S R \, T_1\, R\inv S\inv.  
}
(It should be observed that these expressions are not unique.)

Using the indices of the fixed points, defined in \eqref{indicesFixed}, 
in the sums below, the trace
of a general operator $\cO$ on this orbifold becomes
\equ{
\Tr_{\Tors/G, T, R}\,\Bigl[ \cO \Bigr] = 
\frac 18 \Tr_{\Tors/G, T, R}\,\Bigl[ \cO \Bigr] 
+ \frac 18\cdot \frac 14 \sum_p 
\Tr_{\Fixd_{g\, p}, T{}^\parl_g}\,\Bigl[ R{}^{p}_g \cO_g \Bigr] 
\non \\[2ex] 
+ \frac 18\cdot \frac 1{16} \sum_{p,q}
\Tr_{\Fixd_{g^2\, p,q}, T{}^\parl_{g^2}}\,\Bigl[ R{}^{p,q}_{g^2} 
\cO_{g^2} \Bigr] 
 + \frac 18\cdot \frac 1{4} \sum_{p,q}
\Tr_{\Fixd_{k\, p,q}/\langle g^2 \rangle, T{}^\parl_k}\,\Bigl[ R{}^{p,q}_{k} 
\cO_{k} \Bigr] 
\non \\[2ex]
+ \frac 18\cdot \frac 1{4} \sum_{p,q}
\Tr_{\Fixd_{gk\, p,q}, T{}^\parl_{gk}}\,\Bigl[ R{}^{p,q}_{gk} 
\cO_{gk} \Bigr] 
}
where $\cO_1 = \cO$ and 
\equ{
\arry{ll}{
\cO_{g} = 
\cO\Bigl(x, z; \der_x, \frac{1+i}2 \der_{z_1}, \frac{1-i}2 \der_{z_2} \Bigr),
& 
\cO_{g^2} = 
\cO\Bigl(x, z; \der_x, \frac{1}2 \der_{z_1}, \frac{1}2 \der_{z_2} \Bigr),
\\[2ex]
\cO_{k} = 
\cO\Bigl(x, z; \der_x, \der{}^\parl_{k}, \frac{1}2 \der{}^\perp_{\,k} \Bigr),
&
\cO_{k} = 
\cO\Bigl(x, z; \der_x, \der{}^\parl_{gk}, \frac{1}2 \der{}^\perp_{gk} \Bigr).
}
}
And finally the matrices that appear within the traces at the various
fixed hyper surfaces take the form 
\equ{ 
\arry{l l l}{
R_g^p &= T_{(0_4, p_1, i p_2)}\, R &= 
T_1^{p_1} \, T_{2'}^{p_2}\, R,  
\\[1ex] 
R_{g^2}^{p,q} &= T_{(0_4, p_1 \mbox{+} i q_1, p_2 \mbox{+} i q_2)} 
\, R^2 
&= T_1^{p_1}\, T_{1'}^{q_1}\, T_{2}^{p_2}\, T_{2'}^{q_2}\, R^2, 
\\[1ex]
R_{k}^{p,q} &= T_{(0_4,p+iq, -p-iq)}\, S 
&=  T_1^p \, T_{1'}^q \, T_2^{\mbox{-}p}\, T_{2'}^{\mbox{-}q}\, S, 
\\[1ex]
R_{gk}^{p,q} &= T_{(0_4, p \mbox{+} i q, i p - q)} \, RS
&= T_1^{p}\, T_{1'}^{q}\, T_{2}^{\mbox{-}q}\, T_{2'}^{p}\, RS. 
}
}
Hence we see that even though this is a more complicated orbifold than
the Abelian ones, with the methods described in this work they can be
analyzed in the same fashion, even if we allow for field which are
non--periodic on the torus.

\section{Conclusions}
\labl{sc:concl}

We have discussed a general procedure to compute all kinds of
one--loop amplitudes on a fairly general class of orbifolds. It is
well--known that quantum corrections can be very important in four
dimensions; and even more so in extra dimensions, because these
theories are non--renormalizable. Models based on orbifolds are very
popular to perform investigations of the possible (quantum) physics in
extra dimensions. However, most recent field theory investigations
have been performed with rather simple orbifolds, like
$S^1/\Intr_2$. Therefore, it may not be straightforwardly to extended
such calculations to more extra dimensions or more complicated
orbifold groups. The orbifold group may have subgroups or be
non--Abelian. For this reason it is important to have some general
methods available to perform such field theory calculations on
orbifolds. In this paper we have described the tools, which can be
used to perform such computations. Our main results may be summarized
as follows:

One--loop amplitudes on a general class of orbifolds can be expressed
as a sum of one--loop amplitudes, with trivial orbifolding. Therefore, 
these amplitudes are not more difficult to compute than those on
(lower dimensional) tori. As our results only require that the
corresponding operator is compatible with the orbifolding (i.e.\ it
satisfies condition \eqref{OrbiOperator}) our results
can be applied to a wide range of possible computations on orbifolds.
We have illustrated the general methods exposed in this paper by
consider the examples of the orbifolds: $S^1/\Intr_2$, $T^6/\Intr_4$
and $T^4/D_4$. The first example has been often studied in the past,
and provides the reader with an easy test case of our
general methods, while the other two examples show that they can equally
well be applied to non--prime and even non--Abelian
orbifolds. Moreover, the fields are allowed to have twisted
periodicity conditions around non--contractible cycles of the
torus. This leads to different ``projections'' at the various fixed
points which are then distinguished by the Wilson lines.

In more technical detail our results amount to the following: Any
one--loop calculation can be formulated as the evaluation of a given
operator on the Hilbert space of states, which run around in the loop. The
properties of this Hilbert space are determined by the characteristics
of the vector bundle in which those states take their values. This
more abstract point of view has the advantage, that it allows 
us to treat orbifold calculations for many different types of fields,
corresponding to various bundles, in an integrated way. 
The flat orbifolds, considered in this work, are obtained by taking $d$
dimensional Minkowski spaces divided by an $n$ dimensional lattice
$\gG$, and subsequently by a finite group $G \subset \SO{1,d-1}$. We have
denoted this ``torus'' as $\Tors = \Real^{1,d-1}/\gG$ and the
resulting orbifold by $\Tors/G$. The orbifold group $G$ is often taken
to be Abelian. However, our results apply to Abelian and non--Abelian
orbifolds alike. The properties of fields on this spaces are determined by
their transformation properties under lattice shifts and the action of
the elements of the finite group $G$. The fields need not be
periodic under these lattices shifts, nor do they have to transform
trivially under $G$. Therefore, the bundle in which the
fields live, are described by two homomorphisms $T$ and $R$, encoding
the non--trivial periodicities and orbifold twists, respectively. In
this work we have shown how the traces over the Hilbert spaces twisted
by $R$ can be expressed as a sum of traces over Hilbert spaces which are
not twisted by the orbifolding. These Hilbert spaces may correspond to
bundles over lower dimensional tori or orbifolds, and involve local
projections $R{}^{s_r}_{g_r}$. These local projectors take the effects
of non--trivial Wilson lines into account. In addition, we found that each of
these subtraces come with a normalization factor 
$1/\det{}^\perp_{\,g}(1-g)$, and the derivatives on the delta function
are rescaled, according to 
\(
\der{}^\perp_{\,g} \ra (\der(1-g)\inv){}^\perp_{\,g}. 
\)
We have given a general prescription of how the orbifolded fixed point
spaces can be identified.

Let us close the conclusion by giving some examples to which our
results can be applied. In this work we have mentioned only a few 
applications: the computation of anomalies, tadpoles and vacuum
energies on general orbifolds. However, as the results can be applied to
any operator on an orbifold, it can also be used to compute the
renormalization of the kinetic terms or the (gauge) couplings, both in
the bulk, as well as, on the orbifold fixed hyper surfaces. In this
work we have restricted ourselves to flat global orbifolds, i.e.\ the
covering space has been taken the flat $d$ dimensional Minkowski space
$\Real^{1,d-1}$. We would expect that the methods can be
extended to the case where the covering space is a general curved
manifold, that possesses  a certain amount of isometries. Also, it should be
possible to formulate the prescriptions such that they can be applied
to local orbifolds, where the orbifolding group for the various 
coordinate patches may be different.

\section*{Acknowledgements}

The author would like to thank T.\ Kobayashi and M.G.A.\ Walter, for
inspiring discussions concerning heterotic $T^6/Z_4$ orbifolds that
have initiated this project, and M.\ Olechowski and 
M.\ Vonk for some useful communications. The author also acknowledges 
the kind hospitality of the group of H.P.\ Nilles in Bonn where this
work was completed. 
\\
This work has been supported by NSERC and CITA.

\appendix
\def\theequation{\thesection.\arabic{equation}} 

\setcounter{equation}{0}
\section{Complex scalar functions on $\boldsymbol{\Tors}$}
\labl{sc:scalar}

Complex scalar functions define the simplest examples of sections of
bundles over $\Tors$ or $\Tors/G$. Most other more complicated bundles
can be obtained by taking a tensor product of this complex line bundle
times other bundles. In particular, if one wants to give an explicit
basis for the Hilbert space associated with a given bundle, then one
needs the basis functions of the scalar Hilbert space which we now
describe. The dual or momentum space $\Tors^*$ of $\Tors$ is defined
as  
\equ{
\Tors^* = \{ q \in (\Real^{1,d-1})^* \, |\, 
\forall v \in \gG: ~ q^T\get\, v \in 2\pi\, \Intr \}.  
\labl{DualV}
}
The elements of $\Tors^*$ can be used to parameterize a basis of the
Hilbert space of complex scalar functions on $\Tors$:
\equ{
\gf(x;q) = \frac {e^{i\, q^T \get\, x}}{\sqrt{N}}, 
\quad 
\der_\gm \gf(x; q) = i\, (q^T \get)_\gm \gf(x;q), 
\quad 
\forall v \in \gG:~
\gf(x+v; q) = \gf(x; g). 
\labl{ModeF}
}
which are  periodic w.r.t.\ $\gG$. 
The normalization $N = (2\pi)^{d-n} \text{Vol}_\gG$ takes into account
the usual factors of $2\pi$, that arise with Fourier transforms, and the
volume  $\text{Vol}_\gG$ of a fundamental domain of the lattice $\gG$.
Their orthonormality relations read 
\equ{
\int_{\Tors} \d x\, 
\gf(x; q) \gf(x; q')^* = \gd_{\Tors^*}(q-q'),
\quad 
\int_{\Tors^*} \d q\, \gf(x; q) \gf(x'; q)^* = \gd_{\Tors}(x-x'),
\labl{orthon}
}
Here we have
employed Lebeque integration to avoid having to write sums and
integrals over the discrete and continuous parts of $\Tors^*$
separately. With $\gd_{\Tors^*}(p)$ we denote the Kronnecker delta on
the discrete part of $\Tors^*$ and the ordinary delta on the remaining
$d-n$ directions in $\Tors^*$. 
In section \ref{sc:geom} we defined the projection operators  
$P{}^{\parl,\perp}_{\,g}$ of an element $g \in G$. Because of the
direct product structure \eqref{DirectProd} of $\Tors$, it follows
that the mode functions \eqref{ModeF} factorize as 
\equ{
\gf(x; q) = \gf{}_g^\parl(x;q) \, \gf{}_{\,g}^\perp(x;q),
\labl{DeltaSplitting}
}
Here $\gf_g^\parl(x;q) = \sqrt{N^\perp_{\, g}} \gf(x; P_g^\parl q)$ and 
$\gf_g^\perp(x;q) = \sqrt{N^\parl_g} \gf(x; P_{\,g}^\perp q)$ are
complete sets of mode functions on $\Tors_g^\parl$ and
$\Tors_{\,g}^\perp$, respectively. Here for $i = \parl, \perp$ we have
defined $N^i_g = (2\pi)^{d^i_g-n^i_g} \text{Vol}_{\gG^i_g}$, 
in terms of the dimension $n_g^i$ of the lattice $\gG^i_g$, and its
volume $\text{Vol}_{\gG^i_g}$. 
Hence, we have similar orthonormality relations as \eqref{orthon} on 
$\Tors^{\parl,\perp}_{\,g}$  
\equ{
\int_{\Tors^{i}_g} \d x\, 
\gf{}^{i}_g(x; q) \gf^{i}_g(x; q')^* 
= \gd{}_g^{i}{}^*(q-q'),  
\quad 
\int_{\Tors_g^{i}{}^*} \d q\, 
\gf{}^{i}_g(x; q) \gf^{i}_g(x'; q)^* 
= \gd^{i}_g(x-x'). 
\labl{orthonParlPerp}
}
In addition, a useful property of the perpendicular scalar mode functions is 
\equ{
\gf{}_{\,g}^\perp(gx; q)^* 
\der_x \,  \gf{}_{\,g}^\perp(x; q)  = 
\bigl[ \der_x (1-g)\inv \bigr]{}^\perp_{\,g} 
\gf{}_{\,g}^\perp(x; q) \gf{}_{\,g}^\perp(gx; q)^*. 
\labl{DerOut}
}

Since by assumption the fiber $\Fibr$ is a complex vector space, we
can use these scalar mode functions to define a basis for the Hilbert
space $\cH_{\Tors}$. Let $\ge_\gs$ be a basis for the vector space
$\Fibr$, i.e.\ 
\equ{
\ge_\gs{}^\dag \ge_{\gs'} = \gd_{\gs\, \gs'}
\qquad 
\sum_\gs \ge_\gs \ge_\gs^\dag = \Id. 
\labl{BasisFiber}
}
An orthonormal basis $| \gf_\gs(q) \rangle$ for the Hilbert space
$\cH_{\Tors}$ is given in the coordinate space by 
\(
\gf_\gs(x; q) = \gf(x; q)\, \ge_\gs: 
\)
\equ{
\langle \gf_\gs(q) \,|\, \gf_{\gs'}(q') \rangle = 
\int_\Tors \d x\, \gf_\gs(x; q)^\dag  \gf_{\gs'}(x; q') 
= \gd_{\gs\, \gs'} \gd_{\Tors^*}(q - q').  
\labl{BasisHilbert}
}

\setcounter{equation}{0}
\section{Orbifold Hilbert space trace computations}
\labl{sc:traces}

This appendix is devoted to the derivation of the central results
given in section \ref{sc:OrbiTrace}. In particular we prove the 
identities \eqref{TraceOrbiComp} and \eqref{RemoveOrbiProj}.

To obtain equation \eqref{TraceOrbiComp}, write out the trace in the
orthonormal basis \eqref{BasisHilbert} of the Hilbert space
$\cH_{\Tors,T}$ and the definition of the projection operator
\eqref{OrbiProj}, and subsequently employ the change of variables: 
$y = h\inv x$ and $h\inv g \ra g$:   
\equ{
\Tr_{\Tors, T, R}\,[ \cO ]  = 
\\[2ex]
= \frac 1{|G|^2}  \sum_{g,h\in G} \sum_\gs
\int\limits_{\Tors\otimes \Tors^{*}} \!\! \d x \, \d q~
\gf_\gs(g\inv x;q)^\dag   \cT(g\inv x) \inv  R_g^\dag \, \gr 
\,\cO(x, \der) \,
R_h \cT(h\inv x)  \gf_\gs(h\inv x;q) 
\non 
\\[2ex]
 = \frac 1{|G|^2}  \sum_{g,h \in G} \sum_\gs
\int\limits_{\Tors\otimes \Tors^{*}} \!\! \d y \, \d q~
\gf_\gs({g}\inv y;q)^\dag  \cT({g}\inv y) \inv R_{g}^\dag \, \gr\,  
\,R_{h}\inv \cO(h y, \der_y {h}\inv) R_h\, 
\cT(y)  \gf_\gs(y;q). 
\non 
}
This expression is the explicit form of equation
\eqref{TraceOrbiComp}.

Next we compute the $\Tr_{\Tors,T}\,[\cP_R{}^\dag \cO]$ 
for an arbitrary operator $\cO$ on $\cH_{\Tors,T}$ to confirm 
equation \eqref{RemoveOrbiProj}. We first employ the
splitting of the basis of the Hilbert space into parallel and
perpendicular components w.r.t.\ each $g \in G$ 
\equa{
\Tr_{\Tors, T}\, \Bigl[\cP_R{}^\dag \cO \Bigr] = 
\frac 1{|G|}  \sum_{g \in G} \sum_\gs
& \int\limits_{\Tors\otimes \Tors^{*}} \!\!  \d x \, \d q~ 
\gf^\parl_{g\,\gs}(x;q)^\dag  
\gf^\perp_{\,g}({g}\inv x;q)^*
\cT{}^\parl_g(x)\inv 
\cT{}^\perp_{\,g}({g}\inv x) \inv R_{g}{}^\dag \, \cdot 
\non \\  & 
\cdot \,\gr \,  \sum_{\vec{\gm}} \cO^{\vec{\gm}}(x)\,  \der_{\vec{\gm}}\, 
\cT{}^\perp_{\,g}(x)  \cT{}^\parl_g(x)  
\gf^\perp_{\,g}(x; q) \gf^\parl_{g\,\gs}(x;q). 
}
Here we have used that the points in the subspace $\Tors_g^\parl$ are 
inert under the action of $g$. The next step is to bring all functions
of $x^\perp_{\, g}$ on the right side of the derivatives. To this end, we
move $\cT{}^\perp_{\,g}({g}\inv x) \inv$ to the left, and insert
another decomposition of unity 
$\sum_{\gs'} \ge_{\gs'} \ge_{\gs'}{}^\dag = 1$ after it. Keeping track
on what the derivatives $\der_{\vec{\gm}}$ act, we can take 
$\ge_{\gs}{}^\dag \cT{}^\perp_{\,g}({g}\inv x) \inv \ge_{s'}$ and 
$\gf^\perp_{\,g}({g}\inv x;q)^*$ to the far r.h.s.\ of this equation. 
Using the closure $\sum_{\gs} \ge_{\gs} \ge_{\gs}{}^\dag = 1$ once
again, and the matrix version of \eqref{DerOut} 
\equa{
\der_{\vec{\gm}} \Bigl( 
 \gf{}_{\,g}^\perp(x; q) \cT{}^\perp_{\,g}(x) 
\Bigr) \,& \Bigl[
 \cT{}^\perp_{\,g}({g}\inv x) \inv 
\gf{}_{\,g}^\perp(g\inv x; q)^* \Bigr] 
=  
\\  & 
\Bigl[ \der^\parl_g + (\der (1-g\inv)\inv) {}^\perp_{\,g} 
\Bigr]_{\vec{\gm}}
\Bigl(
\cT{}^\perp_{\,g}(x) \cT{}^\perp_{\,g}({g}\inv x) \inv 
\gf_{\,g}^\perp(x; q) \gf{}_{\,g}^\perp(g\inv x; q)^* 
\Bigr),
\non 
}
we obtain 
\equ{
\Tr_{\Tors, T}\, \Bigl[\cP_R{}^\dag \cO \Bigr] = 
\frac 1{|G|}  \sum_{g \in G} \sum_\gs
\int\limits_{\Tors\otimes \Tors{}^\parl_{g}{}^{*}} \!\!  \d x \, \d q~  
\gf^\parl_{g\,\gs}(x;q)^\dag  
\cT{}^\parl_g(x)\inv 
R_{g}{}^\dag \, \gr \, 
\sum_{\vec{\gm}} \cO^{\vec{\gm}}(x)\,  
\cT{}^\perp_{\,g}(x)
\, \cdot 
\\  
\cdot \,  
\Bigl[ \der{}^\parl_g + (\der (1 - g\inv)\inv){}^\perp_{\,g} 
\Bigr]_{\vec{\gm}}
 \Bigl( 
\cT{}^\parl_g(x)  
\cT{}^\perp_{\,g}({g}\inv x) \inv 
\gf^\parl_{g\,\gs}(x;q)
\int\limits_{\Tors{}^{\perp*}_{\,g}} \!\! \d q~
\gf^\perp_{\,g}({g}\inv x;q)^* \gf^\perp_{\,g}(x; q) 
\Bigr). 
\non
}
The closure relation \eqref{orthonParlPerp} on $\Tors_g^\perp$ and
\eqref{FixedDelta} allows us to rewrite this as contributions on the
fixed spaces of $g$:  
\equ{
\Tr_{\Tors, T}\, \Bigl[\cP_R{}^\dag \cO \Bigr] = 
\frac 1{|G|}  \sum_{g \in G} 
\frac 1{|\det^\perp_g (1\!\!-\!g\inv)|}
\sum_{s,\gs} \!\!\!
\int\limits_{\Tors\otimes \Tors{}^{\parl*}_g} \!\!\!\!  \d x \, \d q~  
\gf^\parl_{g\,\gs}(x;q)^\dag  
\cT{}^\parl_g(x)\inv 
R_{g}{}^\dag \, \gr \,
\cdot 
\labl{InterRes}
\\
\cdot \, 
\sum_{\vec{\gm}} \cO^{\vec{\gm}}(\fZ{}_g^s+x^\parl_g) 
\Bigl[ 
\der{}^\parl_g + (\der (1 \!\!-\!\! g\inv)\inv){}^\perp_{\,g} 
\Bigr]_{\vec{\gm}} \Bigl( 
\cT{}^\perp_{\,g}(\fZ{}^s_g) 
\cT{}^\perp_{\,g}({g}\inv \fZ{}_g^s) \inv 
\cT{}^\parl_g(x)  \gf^\parl_{g\,\gs}(x;q)
\, 
\gd^\perp_{\,g}(x - \fZ{}^s_g)
\Bigr).
\non 
}
Here we have substituted $x^\perp_g = \fZ^s_g$ everywhere, using the 
defining property of the delta function $\gd{}^\perp_g$. It should be
noted, that one has to be careful employing this property here, because
of the presence of the derivative $\der{}^\perp_{g}$. However, since
we made sure in the previous step that all $x^\perp_g$ dependence is
under the derivatives,  we may safely use this property.\footnote{Let
$a$ be any function and $f$ a test function of $x$, we have that 
\(
\int \d x\, \der(a(x) \gd(x)) f(x) = - \int \d x\, a(x) \gd(x) \der
f(x) = - \int \d x\, a(0) \gd(x) \der f(x) = \int \d x\, a(0) \der
(\gd(x)) f(x).
\)} 
Since now,  
$\cT{}^\perp_{\,g}(\fZ^s_g) \cT{}^\perp_{\,g}({g}\inv \fZ^s_g) \inv$
is constant, we can pull it outside of the derivatives. Taking 
$g \ra g\inv$ and using that the projector of $g$ and $g\inv$ are
identical (see \eqref{gProj}), and employing the reserve method of
including complete sets, allows us to rewrite this as 
\equa{
\Tr_{\Tors, T}\, \Bigl[\cP_R {}^\dag \cO \Bigr] = & 
\frac 1{|G|}  \sum_{g \in G} 
\frac 1{|\det^\perp_g (1\!\!-\!g)|}
\sum_\gs
\int\limits_{\Fixd_{\,g} \otimes \Tors{}^{\parl*}_g} \!\!  \d x \, \d q~  
\gf^\parl_{g\,\gs}(x;q)^\dag  
\cT{}^\parl_g(x)\inv 
\cR_g(x)
\, \cdot 
\non  \\  & 
\cdot \,  
\sum_{\vec{\gm}} \cO^{\vec{\gm}}(x)\,  
\Bigl[ 
\der{}^\parl_g +  (\der (1 - g)\inv){}^\perp_{\,g} 
\Bigr]_{\vec{\gm}}
\cT{}^\parl_g(x)  \gf^\parl_{g\,\gs}(x;q), 
}
where we have employed \eqref{Projgv} and \eqref{ProjgvGen}; 
and thereby we have obtained  \eqref{RemoveOrbiProj}.

\bibliographystyle{paper}
{\small
\bibliography{paper}
}

\end{document}